\documentclass[fleqn,usenatbib]{mnras}

\usepackage{newtxtext,newtxmath}
\usepackage[T1]{fontenc}

\usepackage{graphicx}	% Including figure files
\usepackage{amsmath}	% Advanced maths commands
\usepackage{multicol}        % Multi-column entries in tables
\title[galaxy formation and 21-cm signal]{POLAR - I: linking the 21-cm signal from the epoch of reionization to galaxy formation}

\author[Ma et al.]{Qing-Bo Ma$^{1,2}$ \thanks{E-mail: \href{mailto:maqb@gznu.edu.cn}{maqb@gznu.edu.cn}},
Raghunath Ghara$^{3}$,
Benedetta Ciardi$^{4}$,
Ilian T. Iliev$^{5}$,
Léon V. E. Koopmans$^{6}$,
\newauthor Garrelt Mellema$^{7}$,
Rajesh Mondal$^{8}$,
Saleem Zaroubi$^{3,6}$\\
$^{1}$School of Physics and Electronic Science, Guizhou Normal University, Guiyang 550001, PR China \\
$^{2}$Guizhou Provincial Key Laboratory of Radio Astronomy and Data Processing, Guizhou Normal University, Guiyang 550001, PR China \\
$^{3}$ ARCO (Astrophysics Research Center), Department of Natural Sciences, The Open University of Israel, 1 University Road, PO Box 808, Ra'anana 4353701, Israel \\
$^{4}$Max-Planck-Institut f\"ur Astrophysik, Karl-Schwarzschild-Stra\ss e 1, D-85748 Garching bei M\"unchen, Germany\\
$^{5}$Astronomy Centre, Department of Physics and Astronomy, University of Sussex, Falmer, Brighton BN19QH, UK\\
$^{6}$Kapteyn Astronomical Institute, University of Groningen, P.O. Box 800, 9700AV Groningen, the Netherlands\\
$^{7}$The Oskar Klein Centre, Department of Astronomy, Stockholm University, AlbaNova, SE-10691 Stockholm, Sweden\\
$^{8}$Department of Astrophysics, School of Physics and Astronomy, Tel Aviv University, Tel Aviv 69978, Israel
}

\pubyear{2023}

\begin{document}
\label{firstpage}
\pagerange{\pageref{firstpage}--\pageref{lastpage}}
\maketitle

\begin{abstract}
To self-consistently model galactic properties, reionization of the intergalactic medium, and the associated 21-cm signal, we have developed the algorithm {\sc polar} by integrating the one-dimensional radiative transfer code {\sc grizzly} with the semi-analytical galaxy formation code {\sc L-Galaxies 2020}. 
Our proof-of-concept results are consistent with observations of the star formation rate history, UV luminosity function and the CMB Thomson scattering optical depth. 
We then investigate how different galaxy formation models affect UV luminosity functions and 21-cm power spectra, and find that while the former are most sensitive to the parameters describing the merger of halos, the latter have a stronger dependence on the supernovae feedback parameters, and both are affected by the escape fraction model.  
\end{abstract}


%%%%%%%%%%%%%%%%%%%%%%%%%%%%%%%%%%%%%%%%%%%%%%%%%%

%%%%%%%%%%%%%%%%% BODY OF PAPER %%%%%%%%%%%%%%%%%%

%%%%%%%%%%%%%%%%%%%%%%%%%%%%%%%%%%%%%%%%%%%%%%%%%%
\section{Introduction}
\label{sec:intro}
The Epoch of Reionization (EoR) refers to the period when the Universe transitioned from a nearly fully neutral to a highly ionized phase, following the formation of the first galaxies and stars \citep{Furlanetto2006, Dayal2018}.
Observations of the Gunn-Peterson (GP) absorption trough in the spectra of high-$z$ QSOs suggest that the EoR is finished at $z \sim 6$ \cite[e.g.][]{Fan2006}, 
although the long GP troughs detected in the Ly$\alpha$ forest at $z<6$ \cite[e.g. ][]{Becker2015, Bosman2022} indicate a later ending.
The most recent observations of the Cosmic Microwave Background (CMB), e.g. with the {\it Planck} satellite \citep{Planck2020}, have measured a Thomson scattering optical depth $\tau= 0.054\pm0.007$, which implies a mid-point redshift of the EoR (i.e. a global ionization fraction $\bar{x}_{\rm HII} = 0.5$) at $z=7.68\pm0.79$. 
The initial phases of the EoR are still poorly known, although the Experiment to Detect the Global EoR Signature (EDGES) project reported an absorption profile of global 21-cm signal at 78~MHz (i.e. $z\sim 17$) \citep{Bowman2018}, which can be used to put some constraints on the first sources of ionizing radiation. 
Note that this result is still strongly debated \cite[e.g. ][]{Hills2018, Singh2022}, and has not been confirmed by the SARAS 3 project \citep{Bevins2022}.

The galaxies that formed during the EoR are expected to be the main sources of ionization of neutral hydrogen (HI). 
The properties of these $z>6$ galaxies have been studied with hydrodynamical and/or radiative transfer simulations, such as THESAN \citep{Kannan2022}, ASTRID \citep{Bird2022}, CROC \citep{Esmerian2021}, SPHINX \citep{Rosdahl2018} and FIRE \citep{MaX2018},
as well as with more efficient semi-analytical/numerical approaches such as the ReionYuga \citep{Mondal2017}, the ASTRAEUS \citep{Hutter2021} and the MERAXES \citep{Mutch2016, Balu2023} models.
All these simulations predict galactic properties that are generally consistent with high-$z$ observations, e.g. in terms of galaxy stellar mass functions, UV luminosity functions, and star formation history \cite[see the comparisons by e.g. ][]{Kannan2022}. 
With the development of new observational facilities, an increasing number of high-$z$ galaxies have been detected \citep{Stark2016}. 
For example, the Hubble Space Telescope (HST) and the Spitzer telescope have already provided abundant data to build rest-frame UV luminosity functions and stellar mass functions of galaxies at $z>6$ \cite[e.g.][]{Bouwens2021, Stefanon2021a}. 
The Atacama Large Millimeter/sub-millimeter Array (ALMA) telescope has also identified several high-$z$ galaxies through e.g. the [CII] line \citep{Bouwens2020}.
Despite having been collecting data for less than one year, the James Webb Space Telescope (JWST) has already found many new high-$z$ galaxies \cite[e.g.][]{Donnan2022, Harikane2023}, possibly as high as  
$z \sim 17$ \citep{Harikane2023}. 
JWST is expected to observe many more such galaxies in the near future \citep{Steinhardt2021}, thus offering the possibility to massively improve our knowledge of primeval objects.

The UV and X-ray radiation emitted in high-$z$ galaxies, e.g. from stellar sources, X-ray binaries and accreting massive black holes, is expected to change the ionization and temperature state of the HI within the intergalactic medium (IGM) \citep{Islam2019, Eide2020}. 
The radiation emitted through the hyperfine structure transition of high-$z$ HI (with a rest-frame wavelength of $\sim$21-cm) can be measured by modern low-frequency radio facilities \citep{Furlanetto2006}.
Some early results from 21-cm telescopes have put upper limits on the 21-cm power spectra $\Delta_{\rm 21cm}^{2}$ from the EoR, e.g. a 2-$\sigma$ upper limit of $\Delta_{\rm 21cm}^{2} < (73)^{2}\,\rm mK^{2}$ at $k=0.075\,h\,\rm cMpc^{-1}$ and $z\approx9.1$ from the Low-Frequency Array (LOFAR) \citep{Mertens2020}, of $\Delta_{\rm 21cm}^{2} \le (43)^{2}\,\rm mK^{2}$ at $k=0.14\,h\,\rm cMpc^{-1}$ and $z=6.5$ from the Murchison Widefield Array (MWA) \citep{Trott2020}, and of $\Delta_{\rm 21cm}^{2} \le (30.76)^{2}\,\rm mK^{2}$ at $k=0.192\,h\,\rm cMpc^{-1}$ and $z=7.9$ from the Hydrogen Epoch of Reionization Array (HERA)  \citep{Abdurashidova2022}.
These results are already used to rule out some extreme EoR models \citep{Ghara2020, Mondal2020, Ghara2021, Greig2021MWA, Greig2021LOFAR, Abdurashidova2022b}.
While analysis of more data from such facilities will set increasingly tighter upper limits (and possibly also a measurement) of the 21-cm power spectrum, the planned Square Kilometre Array (SKA) is expected to provide also 3-D topological images of the 21-cm signal \citep{Koopmans2015, Mellema2015, ghara16}.

Since both the infrared to sub-mm radiation from high-$z$ galaxies and the 21-cm signal are produced during the EoR, the combination of observations in different frequency bands would provide a deeper understanding of the physical processes at play during the EoR. 
With this idea in mind, some codes have been developed to constrain EoR models with Markov Chain Monte Carlo (MCMC) techniques used in combination with multi-frequency observations, e.g. the semi-numerical model by \cite{Park2019, Park2020} based on 21CMMC \citep{Greig2015} and 21cmFAST \citep{Mesinger2011}, as well as analytical models for 21-cm power spectra and galaxy luminosity functions \cite[e.g. ][]{Zhang2022}. 
While these approaches take advantage of both observations of high-$z$ galaxies (e.g. the UV luminosity functions) and 21-cm power spectra, they do not physically model the properties of galaxies, but estimate the UV luminosity functions and the budget of ionization photons based on the halo mass function model.

In this paper, we describe {\sc polar}, a novel semi-numerical model designed to obtain both the high-$z$ galaxy properties and the 21-cm signal in a fast and robust way, by including the semi-analytical galaxy formation model {\sc L-Galaxies 2020} \citep{Henriques2020} within the one-dimensional radiative transfer code {\sc grizzly} \citep{Ghara2018}, which is an updated version of {\sc bears} \citep{Thomas2009}. 
Since {\sc polar} is fast and thus able to produce a large number of different galaxy and reionization models, we will use it in combination with MCMC techniques and observations of e.g. UV luminosity functions and 21-cm power spectra to provide tighter constraints on both the galaxy and IGM properties. 
In this paper, we introduce the new algorithm and how some selected observables are affected by different choices of the parameters used to describe the formation and evolution of galaxies, as well as the escape of ionizing radiation, while in a companion paper we will extend the formalism to include an MCMC analysis and to constrain the parameters.

The paper is organized as follows: we describe {\sc L-Galaxies 2020} and {\sc grizzly} in Section~\ref{sec:meth}, the resulting galaxy properties and EoR signal are presented in Section~\ref{sec:res}, while a discussion  and the conclusions are found in Section~\ref{sec:conc}.
The cosmological parameters adopted in this paper are the final results of the {\it Planck} project \citep{Planck2020}, i.e. $\Omega_{\Lambda}= 0.685$, $\Omega_{m} = 0.315$, $\Omega_{b} = 0.0493$, $h = 0.674$, $\sigma_{8} = 0.811$ and $n_{s} = 0.965$.

%%%%%%%%%%%%%%%%%%%%%%%%%%%%%%%%%%%%%%%%%%%%%%%%%%
\section{Methods}
\label{sec:meth}
To follow the formation and evolution of galaxies, we combine merger trees from {\it N}-body dark-matter simulations with the semi-analytic model (SAM) {\sc L-Galaxies 2020} \cite[abbreviated as LG20 in the following, ][]{Henriques2020}, while the 1-D radiative transfer (RT) code {\sc grizzly} \citep{ghara15a, Ghara2018} is used to model the gas ionization and 21-cm signal. 
While we refer the readers to the original papers for the details about these tools, in the following section we describe the key aspects that are relevant to this work. 

%%%%%%%%%%%%%%%%%%%%%%%%%%%%%%%%%%%%%%%%%%%%%%%%%%
\subsection{Dark-matter simulations}
\label{meth:simul}
The {\it N}-body dark-matter simulations are run with the {\sc gadget-4} code \citep{Springel2021}, with a box length of $100\, h^{-1}{\rm cMpc}$ and a particle number of $1024^{3}$, i.e. a particle mass $1.2 \times 10^{8}\,{\rm M_{\odot}}$. 
In the following, we will refer to these simulations as L100.
The dark-matter halos are identified with a Friend-of-Friend (FoF) algorithm \citep{Springel2001}, while Subfind is used to identify gravitationally bound sub-halos within halos. 
The merger trees are constructed by following \cite{Springel2005}.
Note that the sub-halos are chosen to have at least 20 dark-matter particles, i.e. the minimum mass is $\sim 2.4\times 10^{9}\,{\rm M_{\odot}}$.
We employ a total of 56 snapshots equally spaced in time in the redshift range $z = 6-20$. 

To resolve the effects of fainter galaxies during the EoR, we also run a smaller simulation with the same $1024^{3}$ particles but box length $35\,{h^{-1}\rm cMpc}$ (abbreviated as L35), able to resolve sub-halos with a minimum mass of $\sim 1.0\times 10^{8}\,{\rm M_{\odot}}$.
\begin{figure}
\centering
    \includegraphics[width=0.95\linewidth]{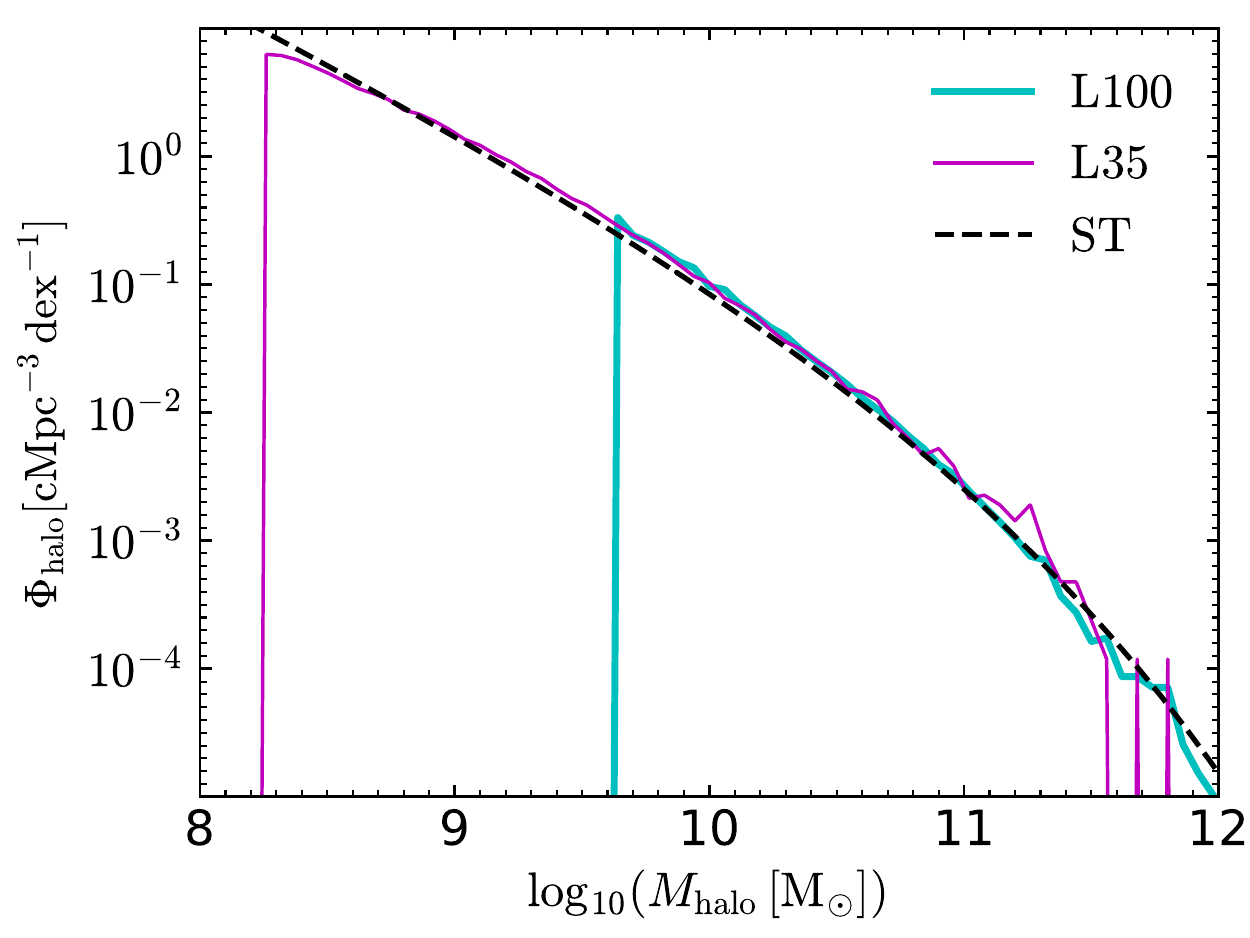}
    \caption{Mass functions of halos ($\Phi_{\rm halo}$) at $z=7$ from simulations L100 (cyan thick line) and L35 (magenta thin line).
    As a reference, the Sheth-Tormen (ST) halo mass function is shown as dashed black line. 
    }
\label{fig:halo_mf_z7}
\end{figure}
As a reference, Fig.~\ref{fig:halo_mf_z7} shows the halo mass functions ($\Phi_{\rm halo}$) at $z=7$ from the two simulations, where $\Phi_{\rm halo} = {\rm d\,}n_{\rm halo}/{\rm d\, log}_{10}(M_{\rm halo})$, with $n_{\rm halo}$ the number density of halos (in units of $\rm cMpc^{-3}$) and $M_{\rm halo}$ halo mass. 
As a reference, we also show the Sheth-Tormen (ST) halo mass function at $z=7$, which is computed with the COLIBRI\footnote{https://github.com/GabrieleParimbelli/COLIBRI} library.
L35 covers a halo mass range of $(1.7 \times 10^{8}-3.6\times10^{11})\,\rm M_{\odot}$, while L100 has halos with mass $(4.2 \times 10^{9}-10^{12})\,\rm M_{\odot}$.
Within the range $(4.2\times 10^{9}-3.6\times 10^{11})\,\rm M_{\odot}$, the halo mass functions of these two simulations are broadly consistent, and both of them are roughly consistent with the ST halo mass function.

%%%%%%%%%%%%%%%%%%%%%%%%%%%%%%%%%%%%%%%%%%%%%%%%%%
\subsection{The semi-analytic code {\sc L-Galaxies 2020}}
\label{meth:sam}
LG20 includes almost all the known physical processes related to galaxy formation \citep{Henriques2020}, e.g. gas cooling, star formation, galaxy merger, supernovae feedback, black hole growth and AGN feedback.
Compared to the previous version \cite[i.e. ][]{Henriques2015}, LG20 adds molecular hydrogen formation, chemical enrichment and spatial tracking of the gas and stellar disc in galaxies, models of stellar population synthesis, dust, tidal effects, and reincorporation of ejected gas.

Specifically, the star formation rate (SFR) is proportional to the H$_{2}$ surface density \citep{Fu2013}, i.e $\Sigma_{\rm SFR} = \alpha_{\rm H_{2}} \Sigma_{\rm H_{2}}/t_{\rm dyn}$, 
where the star formation efficiency $\alpha_{\rm H_{2}}$ is a free parameter, and $t_{\rm dyn}$ is the galactic dynamical timescale as a function of halo mass. 
The H$_{2}$ surface density $\Sigma_{\rm H_{2}}$ is modeled through the cold gas mass, the H$_{2}$ fraction within the H surface density, and the metallicity.

A burst of star formation happens after a halo falls into a larger system, i.e. halo merger, with a time delay $t_{\rm friction}$ due to dynamical friction.
In LG20, $t_{\rm friction}$ is computed with the formulation of \cite{Binney1987}, which depends on the mass and radius of the two merging halos:
\begin{equation}
    t_{\rm friction} = \alpha_{\rm friction} \frac{V_{\rm 200c}r_{\rm sat}^{2}}{GM_{\rm sat,tot}{\rm ln}\Lambda},
\end{equation}
where the efficiency factor $\alpha_{\rm friction}$ is a free parameter, $G$ is the gravitational constant, $r_{\rm sat}$ is the radius of the satellite galaxy, $M_{\rm sat,tot}$ is the sum of dark-matter and baryonic mass of the satellite galaxy, ${\rm ln}\Lambda = {\rm ln}(1+M_{\rm 200c}/M_{\rm sat,tot})$ is the Coulomb logarithm, $M_{\rm 200c}$ and $V_{\rm 200c}$ are the virial mass and velocity of the major halo with overdensity larger than 200 times the critical value of cosmic density. 
The SFR triggered by mergers is modeled through the "collisional starburst" formulation \citep{Somerville2001}:
\begin{equation}
\label{eq:sfr_burst}
    {\rm SFR}_{\rm burst} = \alpha_{\rm SF,\, burst}\left(\frac{M_1}{M_2}\right)^{\beta_{\rm SF,\, burst}} M_{\rm cold,\, tot},
\end{equation}
where $\alpha_{\rm SF,\, burst}$ and $\beta_{\rm SF,\, burst}$ are two free parameters describing the star formation efficiency of a burst, $M_1$ and $M_2$ are the baryonic mass of the two merging galaxies with $M_1 < M_2$, and $M_{\rm cold, tot}$ is their total cold gas mass. 

Supernovae explosions happen at the end of the stellar lifetime, reheating the cold gas and enriching the ISM with metals.
In LG20, the mass reheated by supernovae is proportional to the stellar mass returned into the ISM ($\Delta M_{\star}$), i.e. $\Delta M_{\rm reheat} = \epsilon_{\rm disk} \Delta M_{\star}$, where $\epsilon_{\rm disk}$ is the efficiency factor given by \citep{Henriques2020}:
\begin{equation}
\label{eq:novae_feedback}
    \epsilon_{\rm disk} = \epsilon_{\rm reheat} \times \left[ 0.5+\left(\frac{V_{\rm max}}{V_{\rm reheat}}\right)^{-\beta_{\rm reheat}} \right],
\end{equation}
where $\epsilon_{\rm reheat}$, $V_{\rm reheat}$ and $\beta_{\rm reheat}$ are three free parameters.
$V_{\rm max}$ is the maximum circular velocity of the dark-matter halo, which is related to the halo mass. 
Note that the energy required to heat such mass, i.e. $\Delta E_{\rm reheat} = \frac{1}{2} \Delta M_{\rm reheat} V_{\rm 200c}^{2}$, should be lower than the energy $\Delta E_{\rm SN}$ released by supernovae that is effectively available to the gas components.
Since halos at $z>6$ are generally not very massive, we assume that the condition $\Delta E_{\rm reheat} < \Delta E_{\rm SN}$ is always satisfied.

There are two channels to grow the mass of massive black holes within galaxies \citep{Croton2006}.
The main channel is the halo merger, that can trigger a strong accretion of the central black holes (i.e. quasar mode).
The accreted gas mass of the merger between two neighbouring snapshots (with time difference $t_{\rm diff}$) depends on the properties of the two galaxies \citep{Henriques2015}:
\begin{equation}
    \Delta M_{\rm BH,Q} = \frac{f_{\rm BH}M_{\rm cold, tot}\times(M_{\rm sat}/M_{\rm cen})}{1+(V_{\rm BH}/V_{\rm 200c})^{2}},
\end{equation}
where $M_{\rm cen}$ and $M_{\rm sat}$ are the baryon masses of the central galaxy and satellite galaxy, the fraction of accreted cold gas into black hole $f_{\rm BH}$ and the virial velocity $V_{\rm BH}$ at which the accretion saturates are two free parameters.
The accretion rate can be simply estimated as $\Delta M_{\rm BH,Q}/ t_{\rm diff}$, while the actual accretion rate might be higher, as $t_{\rm diff}$ might be larger than the real lifetime of the quasar. 
The other channel is the accretion of hot gas (i.e. radio mode), which is also the main source of the AGN feedback on star formation.
Its accretion rate is computed with a modified version of the model proposed by \cite{Croton2006}:
\begin{equation}
\label{eq:bh_accre}
    \dot{M}_{\rm BH} = k_{\rm AGN} \left(\frac{M_{\rm hot}}{10^{11}{\rm M}_{\odot}}\right) \left(\frac{M_{\rm BH}}{10^{8}{\rm M_{\odot}}}\right),
\end{equation}
where the accretion efficiency $k_{\rm AGN}$ is a free parameter, $M_{\rm hot}$ is the mass of hot gas within the host galaxy, and $M_{\rm BH}$ is the black hole mass.

In this work, we only focus on the star formation efficiency, halo merger, supernovae feedback and AGN feedback, and keep the default models for other processes, e.g. the gas cooling, the chemical enrichment, the reincorporation of ejected gas, the tidal and ram-pressure stripping and the tidal disruption.
We do not apply the dust model of LG20, but assume the escape fraction to compute the UV luminosity function and the budget of ionization photons following \cite{Park2019}:
\begin{equation}
\label{eq:es_frac}
    f_{\rm es, \lambda} = f_{0,\lambda}\left(\frac{M_{\rm star}}{10^{8}\,\rm M_{\odot}}\right)^{\beta_{\rm es}},
\end{equation}
where $f_{0,\lambda}$ is a function of the photon wavelength $\lambda$ in rest-frame, $M_{\rm star}$ is the stellar mass within the galaxy, and the index factor $\beta_{\rm es}$ is a free parameter to describe the dependence of $f_{\rm es, \lambda}$ on the stellar mass of the galaxy.
Note that $f_{0,\lambda}$ includes the dependence of absorption of dust and neutral gas on the frequency of the emitted photons,
so that its value should be lower for H ionizing photons than for non-ionizing photons, as in the latter case only absorption from dust is effective.
To simplify the discussion, we use only two values for $f_{0, \lambda}$, i.e. $f_{0, \lambda} = 0.25$ at $\lambda = 1600$ \AA~(to match the UV luminosity function of our fiducial model with observations at $z=7$; see Fig.~\ref{fig:lum_uv_dis_z7}), and $f_{0, \lambda} = 0.1$ for ionizing photons (in order for our fiducial reionization history to be consistent with the Thomson scattering optical depth measured by CMB experiments; see discussion in Sec.~\ref{res:21cm}).
In the following, therefore, only $\beta_{\rm es}$ is a free parameter.

\begin{table*}
    \centering
    \begin{tabular}{c c c c c c c c c c c c c c c}
    \hline
             &   $\alpha_{\rm H_{2}}$ & $\alpha_{\rm friction}$ & $\alpha_{\rm SF,burst}$ & $\beta_{\rm SF, burst}$ & $\epsilon_{\rm reheat}$ & $V_{\rm reheat}$ & $\beta_{\rm reheat}$ & $f_{\rm BH}$ & $V_{\rm BH}$ & $k_{\rm AGN}$ & $\beta_{\rm es}$\\
    \hline
    fiducial & 0.06  & 1.8  & 0.5 & 0.38  & 5.6  & 110 & 2.9  & 0.066  & 700  & 0.0025 & 0\\
    smaller & 0.012 & 0.36 & 0.1 & 0.076 & 1.12 & 22  & 0.58 & 0.0132 & 140  & 0.0005 & -0.5\\
    larger  & 0.3   & 9    & 2.5 & 1.9   & 28   & 550 & 14.5 & 0.33   & 3500 & 0.0125 & 0.5\\
    \hline
    \end{tabular}
    \caption{Galaxy formation model parameters with their fiducial values from LG20 (see text for details). The smaller one refers to 20\% of the fiducial value, and the larger one to 5 times the fiducial value. Differently, the escape fraction model parameter $\beta_{\rm es}$ has a fiducial value of 0, while the smaller/larger is -0.5/0.5.}
    \label{tab:param_list}
\end{table*}
In summary, in the models considered here, we have 11 free parameters, which are summarized in  Table~\ref{tab:param_list}. 
In Section~\ref{sec:res}, we will investigate how these parameters affect the global SFR history, the UV luminosity function, the reionization history and the 21-cm power spectrum during the EoR.

%%%%%%%%%%%%%%%%%%%%%%%%%%%%%%%%%%%%%%%%%%%%%%%%%%
\subsection{The radiative transfer code {\sc grizzly}}
\label{meth:rt}
Since the above simulations and formalism do not include radiative transfer, which is crucial to properly model the EoR, we use the results of the {\it N}-body dark-matter simulations and LG20 as input for the 1-D radiative transfer code {\sc grizzly} to describe the HI ionization and heating.
{\sc grizzly} is very efficient in evaluating the ionization and heating processes and the differential brightness temperature of the 21-cm signal ($\delta T_{\rm 21cm}$). 
The algorithm is based on pre-computed ionization and temperature profiles of gas for different source and density properties at various redshifts. 
During the later stages of the EoR, when the ionized bubbles merge into bigger ones, {\sc grizzly} also corrects for the effects of overlap by conserving the ionizing budget.

We use the gridded density fields derived from the {\it N}-body simulations and the galactic properties (i.e. stellar mass and stellar age, see below) computed from LG20, as inputs for {\sc grizzly}.
Note that the gas density is assumed to scale constantly with the dark-matter. 
The matter density and galactic properties from the simulations L100 and L35 are gridded with $100^3$ and $35^3$ cells, respectively, ensuring the same cell resolution of $1\,{h^{-1}\rm cMpc}$ for the RT calculation.

The spectral energy distributions (SEDs) of stellar sources are calculated using the Binary Population and Spectral Synthesis (BPASS) code \citep{Stanway2018}. 
To take into account the history of star formation in the evaluation of the physical properties of the ionized regions, we integrate the SED over the stellar age, i.e. the time from the birth of stars to the output redshifts.
We refer to this as an integrated SED (iSED).
Although LG20 can output iSEDs for each galaxy, for convenience, we adopt the one obtained by averaging the iSEDs normalized by the stellar mass of galaxies. 
In this case, the outputs from LG20 required to run {\sc grizzly} are only the stellar mass and the stellar age of galaxies, but not the full SED for each galaxy, that saves computing time both for LG20 and {\sc grizzly}. 
\begin{figure}
\centering
    \includegraphics[width=0.95\linewidth]{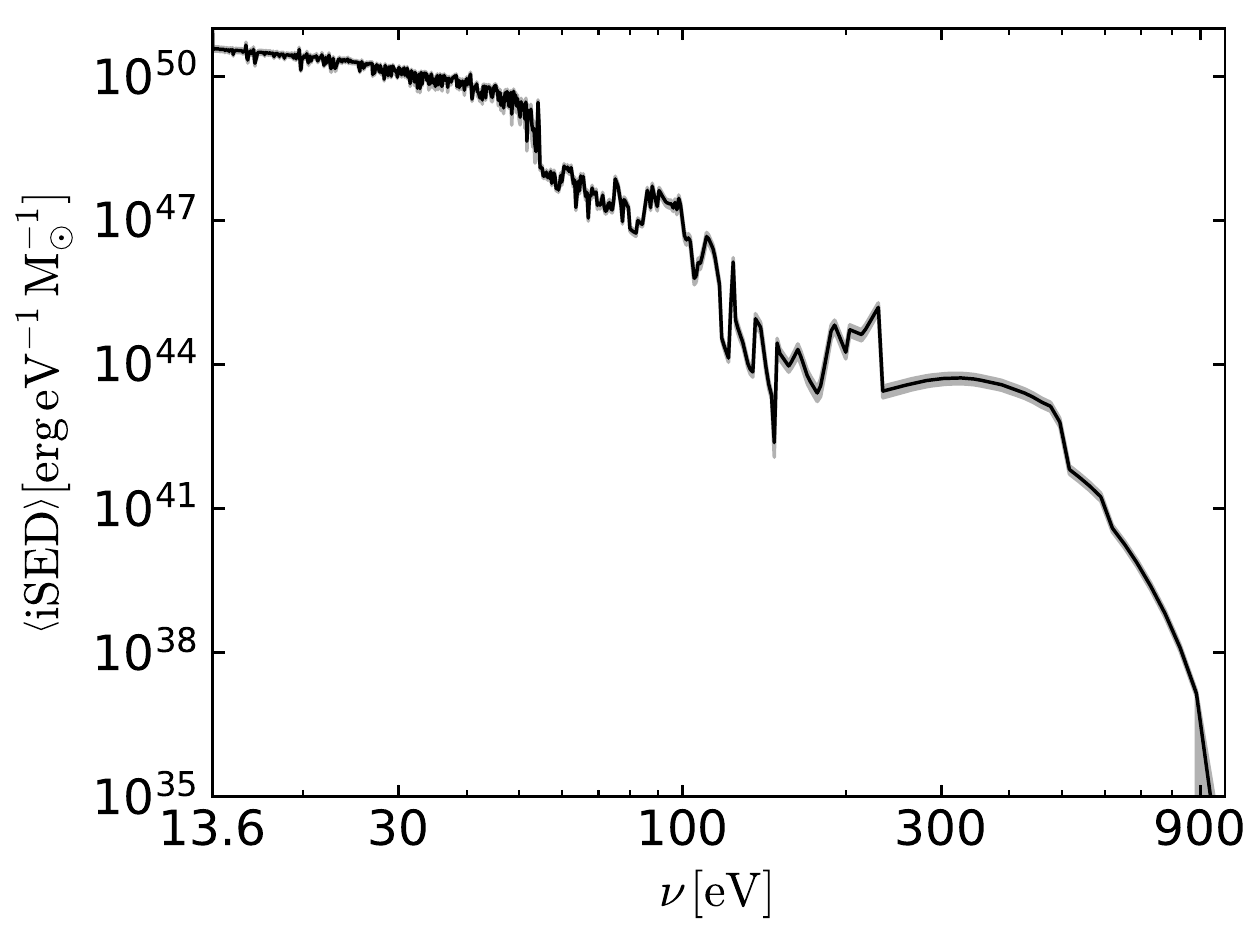}
    \caption{Average iSED ($\rm \left<iSED\right>$) after normalization by the stellar mass of the galaxies (black), with the 1-$\sigma$ area (gray) of $\sim 2.8\times 10^{5}$ galaxies at $z=7$ in the simulation L100 with fiducial parameter values. 
    }
\label{fig:sed_dis_z7}
\end{figure}
As a reference, in Fig.~\ref{fig:sed_dis_z7} we present the average iSED after normalization by the stellar mass of the galaxies with the 1-$\sigma$ area of $\sim 2.8\times 10^{5}$ galaxies at $z=7$, obtained from the simulation L100 with fiducial parameter values. 
This shows that the stellar mass normalized iSEDs have a very small scatter (e.g. root mean square $\sigma$ value is $<10\%$ of the mean value).
We also check that the stellar mass normalized iSEDs are sensitive neither to the galaxy models (i.e. the parameters listed in Table~\ref{tab:param_list} except $\beta_{\rm es}$) nor the output redshifts (see the discussion in Appendix~\ref{app:sed_dep}).
This is due to the fact that, although the UV emission of stellar sources is dominated by massive young stars, both the stellar mass and the iSEDs are integrated over the whole star formation history, i.e. the iSEDs are proportional to the stellar mass of galaxies. 

Finally, {\sc grizzly} computes the $\delta T_{\rm 21cm}$ and the associated power spectrum, where $\delta T_{\rm 21cm}$ is defined as:
\begin{equation}
\label{eq:21cm}
    \delta T_{\rm 21cm} = 27\,{\rm mK}~ \frac{\Omega_{b}h^{2}}{0.023} ~\left(\frac{0.14}{\Omega_{m}h^{2}}\frac{1+z}{10}\right)^{0.5} \times (1 + \delta_{m}) (1 - x_{\rm HII}),
\end{equation}
where $x_{\rm HII}$ is the ionization fraction and $\delta_{m}$ is the over-density of matter. 
As the main goal of this work is to introduce {\sc polar}, we neglect the effect of redshift space distortions, as well as the contribution of X-ray sources, and we assume that the spin temperature ($T_{\rm S}$) is always much larger than the CMB temperature, i.e. $T_{\rm S} \gg T_{\rm CMB}$, focusing instead on the effect of different galaxy formation models.
The power spectrum is estimated as $P_{\rm 21cm}(k) = \left< \delta T_{\rm 21cm}(\vec{k})  \delta T_{\rm 21cm}(-\vec{k}) \right>$, where $\delta T_{\rm 21cm}(\vec{k})$ is the Fourier transfer of $\delta T_{\rm 21cm}$. 
We will use its dimensionless form $\Delta^2_{\rm 21cm}(k) = k^{3}/(2\pi^{2}) \times P_{\rm 21cm}(k)$ in the following discussions.

%%%%%%%%%%%%%%%%%%%%%%%%%%%%%%%%%%%%%%%%%%%%%%%%%%
\section{Results}
\label{sec:res}

To test how the parameters of galaxy formation and escape fraction models affect the observed galaxy properties and 21-cm statistics, we take three values for each parameter: the fiducial ones for the galaxy formation model correspond to the best-fit values by LG20 \citep{Henriques2020}, while the smaller (larger) ones are 20\% (5 times) the fiducial values. The fiducial, smaller and larger value of the escape fraction parameter $\beta_{\rm es}$ is 0, -0.5 and 0.5, respectively. 
All these values are listed in Table~\ref{tab:param_list}.

We note that running {\sc polar} for 56 outputs takes less than two CPU hours, i.e. only a few minutes for each output, with the exact running time depending on the parameter values and the redshift (outputs at lower $z$ are typically more computationally expensive).

%%%%%%%%%%%%%%%%%%%%%%%%%%%%%%%%%%%%%%%%%%%%%%%%%%
\subsection{Properties of galaxies}
\label{res:gala}

\begin{figure*}
\centering
    \includegraphics[width=0.95\linewidth]{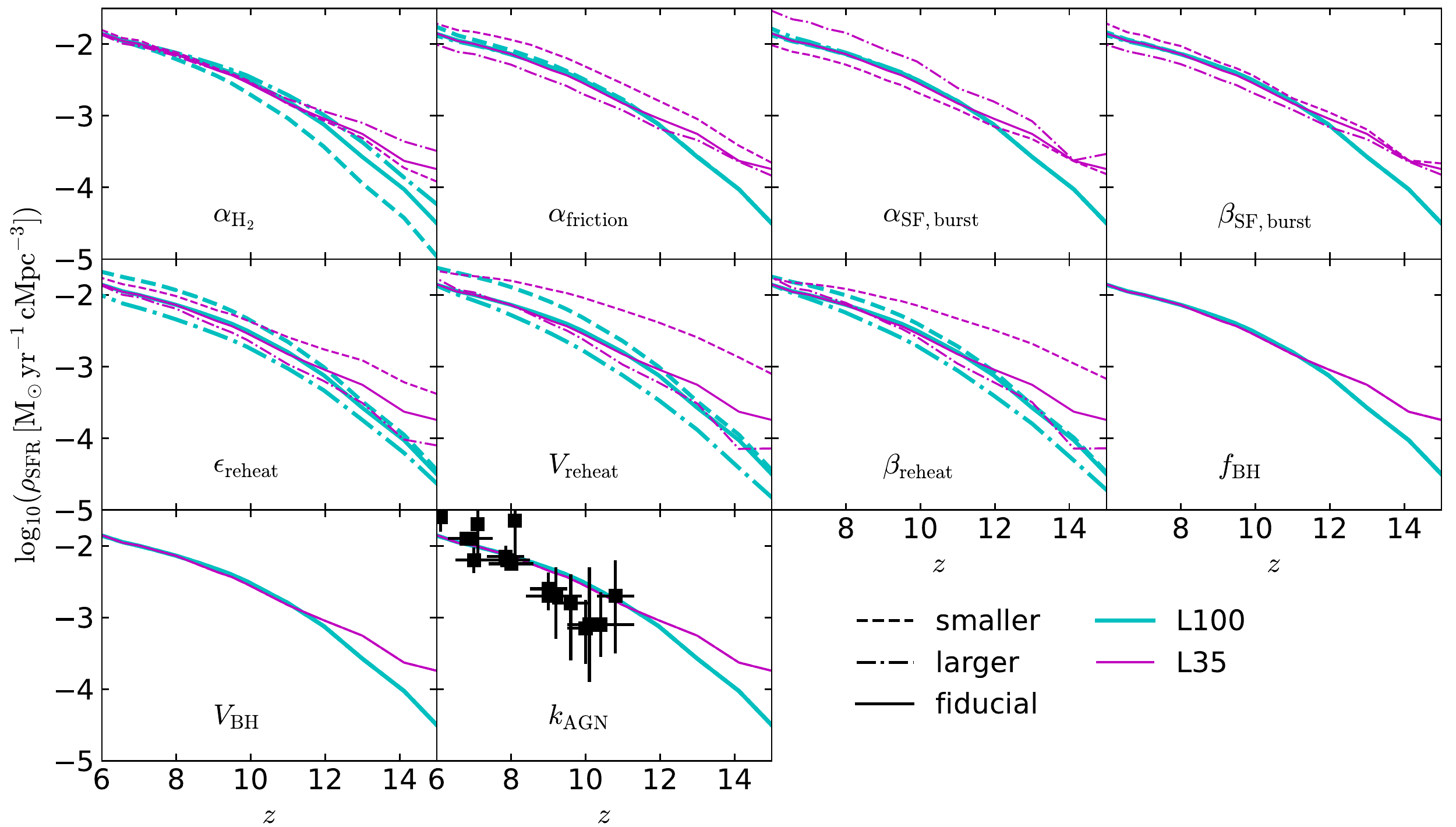}
    \caption{Evolution of SFR density $\rho_{\rm SFR}$ as a function of redshift $z$, for different values of the galaxy formation parameters $\alpha_{\rm H_{2}}$, $\alpha_{\rm friction}$, $\alpha_{\rm SF,\, burst}$, $\beta_{\rm SF,\, burst}$, $\epsilon_{\rm reheat}$, $V_{\rm reheat}$, $\beta_{\rm reheat}$, $f_{\rm BH}$, $V_{\rm BH}$ and $k_{\rm AGN}$, from left to right and from top to bottom. 
    Each panel shows the results of the corresponding parameter with smaller value (dashed line), fiducial value (solid line) and larger value (dash-dotted line) in simulation L100 (cyan thick lines) and L35 (magenta thin lines). 
    The black data points with error-bars refer to the observational data as  summarized in \citet{Ma2017}.
    }
\label{fig:sfr_evol}
\end{figure*}
Fig.~\ref{fig:sfr_evol} shows the evolution of the SFR density $\rho_{\rm SFR}$ as a function of redshift for the 10 parameters (except $\beta_{\rm es}$) describing galaxy formation and evolution as listed in Table~\ref{tab:param_list}. 
As a supplement, in Appendix~\ref{app:stellar_mass} we also present the evolution of the stellar mass density $\rho_{M_{\ast}}$, which presents features similar to those of the $\rho_{\rm SFR}$ shown here. 

\begin{figure*}
\centering
    \includegraphics[width=0.95\linewidth]{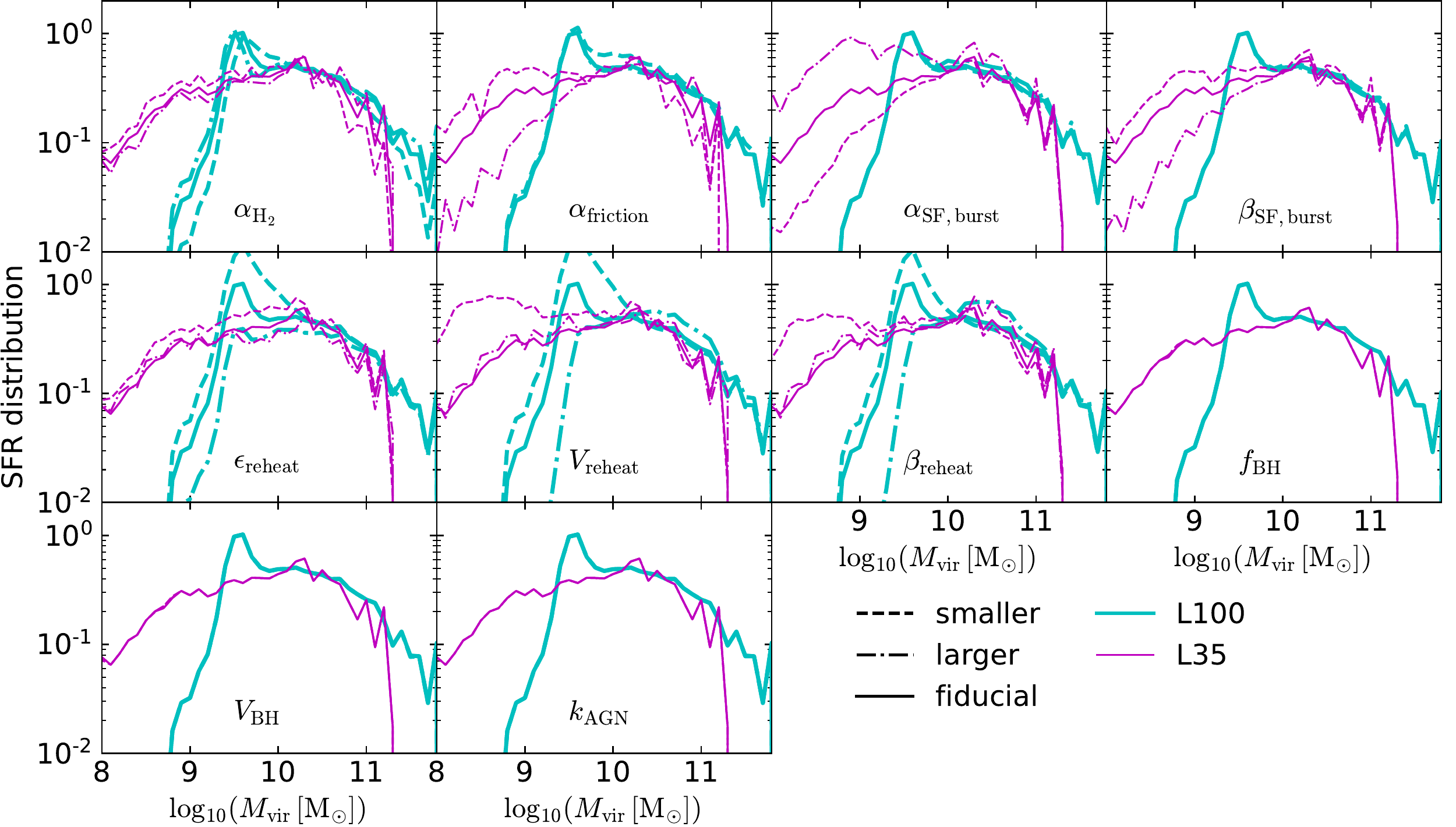}
    \caption{SFR distributions at $z=7$ as functions of the halo virial mass $M_{\rm vir}$, for different values of the galaxy formation parameters $\alpha_{\rm H_{2}}$, $\alpha_{\rm friction}$, $\alpha_{\rm SF,\, burst}$, $\beta_{\rm SF,\, burst}$, $\epsilon_{\rm reheat}$, $V_{\rm reheat}$, $\beta_{\rm reheat}$, $f_{\rm BH}$, $V_{\rm BH}$ and $k_{\rm AGN}$, from left to right and from top to bottom. 
    Each panel shows the results of the corresponding parameter with smaller value (dashed line), fiducial value (solid line) and larger value (dash-dotted line) in simulation L100 (cyan thick lines) and L35 (magenta thin lines). 
    }
\label{fig:sfr_halo_dis_z7}
\end{figure*}

With smaller (larger) star formation efficiency $\alpha_{\rm H_{2}}$, the $\rho_{\rm SFR}$ are as expected lower (higher) at $z > 10$, while they converge at $z<8$ due to supernova feedback effects, i.e. higher star formation results in more supernova feedback that further reduces the star formation (see discussion of Fig.~\ref{fig:sfr_halo_dis_z7} in the following).
The two series of simulations show similar evolution features, but L35 has a higher $\rho_{\rm SFR}$ at $z>12$, as the simulation with higher resolution can resolve more small halos that dominate star formation at such high $z$. 

Changing the merger and starburst parameters (i.e. $\alpha_{\rm friction}$, $\alpha_{\rm SF,\, burst}$ and $\beta_{\rm SF,\, burst}$) has negligible effects on $\rho_{\rm SFR}$ of L100. 
Differently, it changes the results of the higher resolution simulation L35, e.g. the SFR becomes lower by increasing $\alpha_{\rm friction}$ (i.e. higher time delay of mergers), while it increases by increasing the starburst efficiency $\alpha_{\rm SF,\, burst}$. 
Since in Eq.~\ref{eq:sfr_burst} $M_1 < M_2$, an increase of $\beta_{\rm SF,\, burst}$ results in a lower SFR.
As discussed in the following Fig.~\ref{fig:sfr_halo_dis_z7}, this is because the high resolution {\it N}-body simulation resolves more neighboring halos that can more easily merge. 

The supernova feedback parameters (i.e. $\epsilon_{\rm reheat}$, $V_{\rm reheat}$ and $\beta_{\rm reheat}$) affect $\rho_{\rm SFR}$ in both series of simulations. 
A smaller (larger) feedback efficiency (i.e. $\epsilon_{\rm reheat}$) leads to a higher (lower) $\rho_{\rm SFR}$ in both simulations. 
While in L100 the impact is more significant at $z<10$, 
in L35 $\rho_{\rm SFR}$ converges at $z \sim 6$, as star formation here is dominated by merger induced starbursts, thus reducing the effect of supernova feedback on the total SFR.
The parameters $V_{\rm reheat}$ and $\beta_{\rm reheat}$ regulate the dependence of supernovae feedback on the halo mass (see the Eq.~\ref{eq:novae_feedback} and Fig.~\ref{fig:sfr_halo_dis_z7}), which results in different evolution features of $\rho_{\rm SFR}$ in simulations L100 and L35, specifically the latter shows much larger differences than the former at $z>10$.

The AGN parameters (i.e. $f_{\rm BH}$, $V_{\rm BH}$ and $k_{\rm AGN}$) do not have large effects on $\rho_{\rm SFR}$ in both series of simulations.
It is because the energy of AGN is proportional to the black hole mass and the hot gas mass in the galaxies (see Eq.~\ref{eq:bh_accre}), i.e. the AGN feedback is more significant in the most massive halos, which are very rare during the EoR. 

As a reference, we also display observational data as summarized in \citet{Ma2017}.
These are roughly consistent with our predicted $\rho_{\rm SFR}$ from both series of simulations, although the observations still have large error-bars.

To better understand some of the features emerging in Fig.~\ref{fig:sfr_evol}, in Fig.~\ref{fig:sfr_halo_dis_z7} we present the SFR distributions at $z=7$ as functions of the halo virial mass $M_{\rm vir}$. They are computed as $\Delta_{\rm SFR}/\Delta_{{\rm log_{10}}(M_{\rm vir})}/\sum_{\rm SFR}$, where $\Delta_{{\rm log_{10}}(M_{\rm vir})}$ is the bin-width of $M_{\rm vir}$, $\Delta_{\rm SFR}$ is the sum of the SFRs of halos within the bin, and $\sum_{\rm SFR}$ denotes the total SFR. Note that, to have a consistent comparison, the $\sum_{\rm SFR}$ used here is the same for all lines, i.e. from the L100 simulation with the fiducial parameter values.
A smaller (larger) $\alpha_{\rm H_{2}}$ results in a lower (higher) SFR in massive halos (i.e. $M_{\rm vir} >10^{10}\,\rm M_{\odot}$), while the trend is reversed in the less massive ones, especially in the L35 simulation. 
This is due to the effects of supernovae feedback on star formation, i.e. supernovae formed at early times can reduce the SFR in the less massive halos by reheating the cold gas, while this effect is weaker in the massive halos that have higher cooling rates and more cold gas.
Note that supernovae explosions happen with a time delay after the formation of stars, which is very short for massive stars.

The merger and starburst parameters do not affect the SFR distributions as a function of $M_{\rm vir}$ in simulation L100, while they significantly change the SFR of halos with $M_{\rm vir} <10^{10}\,\rm M_{\odot}$ in simulation L35, e.g. the distribution amplitudes with the smaller and larger $\alpha_{\rm SF,\, burst}$ values have $\sim 1$ dex difference at $M_{\rm vir} \sim 10^{8.8}\,\rm M_{\odot}$. 
Since more small halos are resolved in the high resolution simulation (i.e. L35), and they are close to each other, mergers happen more often than in simulation L100. 
However, the SFR within halos with $M_{\rm vir} >10^{10}\,\rm M_{\odot}$ is not very sensitive to the merger models for either simulation.  

A smaller (larger) supernovae feedback efficiency factor $\epsilon_{\rm reheat}$ increases (decreases) the SFR within halos with $M_{\rm vir} <10^{10}\,\rm M_{\odot}$ in simulation L100, while it has no such obvious effect in L35.
One reason is that the star formation of less massive halos in L35 is dominated by mergers, which reduces the impact of supernovae feedback on the SFR. 
As mentioned before, $V_{\rm reheat}$ and $\beta_{\rm reheat}$ relate the supernovae feedback to the halo mass, and thus shape the dependence of SFR on $M_{\rm vir}$. 
For example, in L35, with smaller $V_{\rm reheat}$ and $\beta_{\rm reheat}$ values the SFR of less massive halos ($M_{\rm vir} <10^{9}\,\rm M_{\odot}$) is much higher than the corresponding one with fiducial and larger values, and the latter two cases show similar SFRs.
This is because with smaller $V_{\rm reheat}$ and $\beta_{\rm reheat}$ values the supernovae feedback within less massive halos are much smaller (see Eq.~\ref{eq:novae_feedback}), with the consequence of significantly increasing the SFR of these halos. 
Since the SFR of less massive halos in L35 is dominated by mergers, these overcome the effect of supernovae explosions, so that a higher supernovae feedback (i.e. larger $V_{\rm reheat}$ and $\beta_{\rm reheat}$) does not visibly reduce the SFR. 
At $M_{\rm vir} >10^{9}\,\rm M_{\odot}$ the SFR is similar for all values of  $V_{\rm reheat}$ and $\beta_{\rm reheat}$, i.e. the supernovae feedback effect is weak on SFR within massive halos. 
In L100, the SFRs with smaller $V_{\rm reheat}$ and $\beta_{\rm reheat}$ are higher than those with fiducial values at $M_{\rm vir} <10^{10}\,\rm M_{\odot}$, while they are similar within more massive halos. 
With larger parameter values, the SFRs are lower than those with fiducial values at $M_{\rm vir} <10^{10}\,\rm M_{\odot}$, while they also increase the SFRs of some more massive halos. 

Consistent with earlier results, the effects of the AGN model are only important for the very massive halos, that are rare in our simulations.
Changing all three related parameters does not obviously affect the SFR distributions in the halos of either series of simulations.

\begin{figure*}
\centering
    \includegraphics[width=0.95\linewidth]{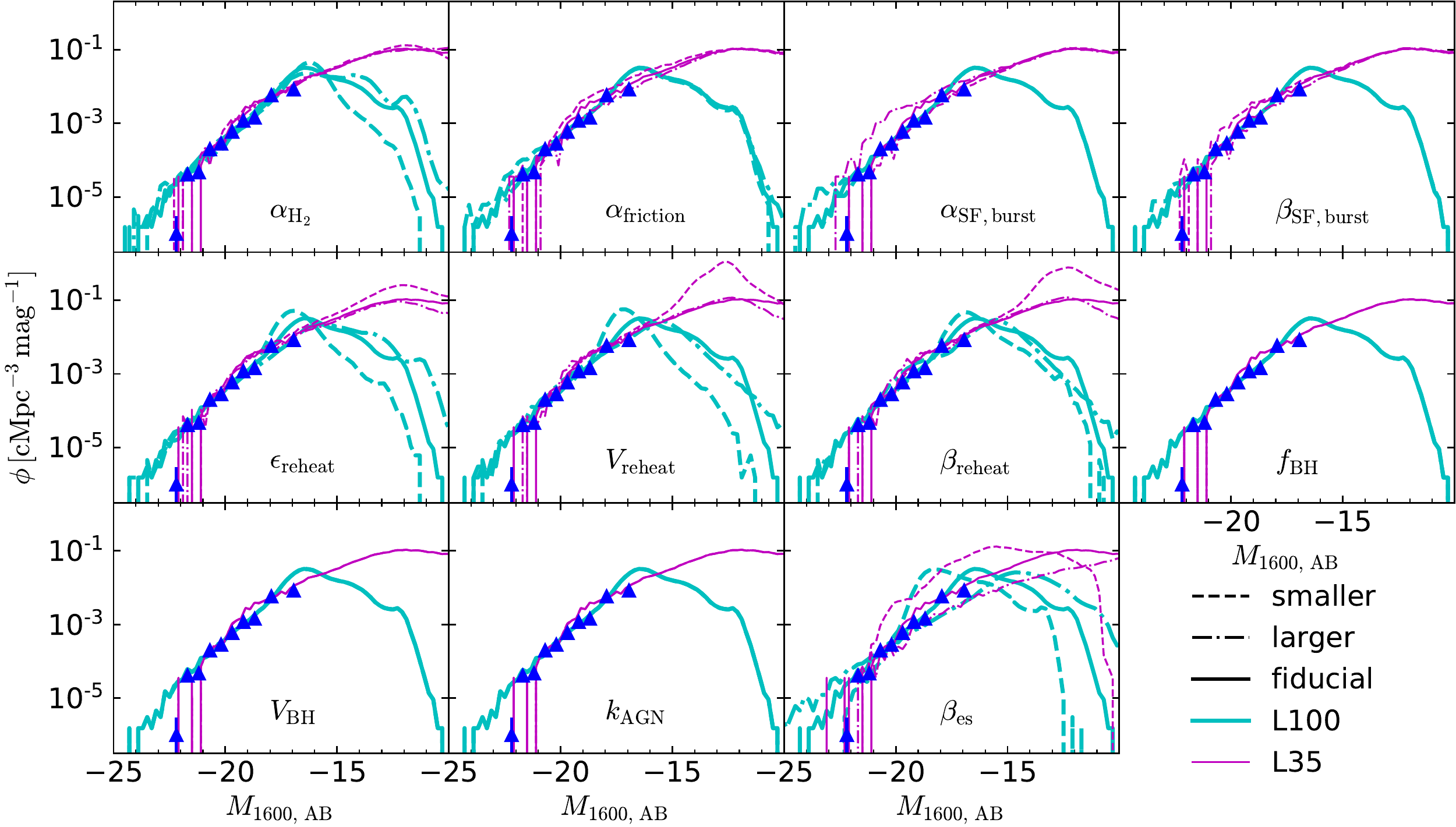}
    \caption{UV luminosity function $\phi$ at the rest-frame wavelength $\lambda = 1600$ \AA~and $z=7$ for different values of the galaxy formation parameters $\alpha_{\rm H_{2}}$, $\alpha_{\rm friction}$, $\alpha_{\rm SF,\, burst}$, $\beta_{\rm SF,\, burst}$, $\epsilon_{\rm reheat}$, $V_{\rm reheat}$, $\beta_{\rm reheat}$, $f_{\rm BH}$, $V_{\rm BH}$, $k_{\rm AGN}$ and $\beta_{\rm es}$, from left to right and from top to bottom. 
    Each panel shows the results of the corresponding parameter with smaller value (dashed line), fiducial value (solid line) and larger value (dash-dotted line) in the simulation L100 (cyan thick lines) and L35 (magenta thin lines). 
    The blue triangle with error-bars are observations at $z=7$ from \citet{Bouwens2021}.
    }
\label{fig:lum_uv_dis_z7}
\end{figure*}
Fig.~\ref{fig:lum_uv_dis_z7} shows the UV luminosity function $\phi$ at the rest-frame wavelength $\lambda = 1600$\AA~for different galaxy formation and escape fraction model parameters. 
The absolute magnitude of galaxy luminosity at $\lambda = 1600$\AA~is computed as:
\begin{equation}
    M_{1600,\, \rm AB} = -\frac{5}{2}{\rm log}_{10}\left(f_{\rm es,\lambda} \frac{F_{1600}}{4\pi R^{2}}\right) -48.6
\end{equation}
where $F_{1600}$ is the galaxy brightness at $\lambda = 1600$\AA, and $R=10\,\rm pc$.
As a comparison, we also show the observations of $\phi$ at $z=7$ \citep{Bouwens2021} (see Appendix~\ref{app:uv_lum} for more comparisons at different redshifts).
Note that, as mentioned earlier, we fix the parameter $f_{0, \lambda} = 0.25$ at $\lambda = 1600$\AA~to match our results with observations. 
The shape and amplitude of the measured $\phi$ are consistent with our fiducial model in simulation L35, while they are slightly lower than the $\phi$ at $M_{1600,\rm \, AB} < -22$ from simulation L100. 
This may be caused by the bias associated to the small field of view of surveys \cite[see e.g. ][]{Bouwens2021}, which might not cover enough bright objects. 

In general, the differences induced on $\phi$ by different parameters are much smaller than those on the SFR distributions shown in Fig.~\ref{fig:sfr_halo_dis_z7}. 
Specifically, three star formation efficiency $\alpha_{\rm H_{2}}$ values result in similar $\phi$ for both simulations, except that the smaller (larger) $\alpha_{\rm H_{2}}$ produces a lower (higher) $\phi$ at $M_{1600,\rm \, AB}>-15$ in L100. 
We note that, limited by the resolution of {\it N}-body simulations, the luminosity functions are not robust at the faint end, due to the lack of low mass halos (see Fig.~\ref{fig:halo_mf_z7}). 

Although the merger parameters $\alpha_{\rm friction}$, $\alpha_{\rm SF,\, burst}$ and $\beta_{\rm SF,\, burst}$ obviously affect the SFR density (Fig.~\ref{fig:sfr_evol}) and the SFR distribution (Fig.~\ref{fig:sfr_halo_dis_z7}) of simulation L35, their effects on $\phi$ are not very significant.
The slight differences are mostly at the bright end (i.e. $M_{1600,\rm \, AB}<-18$).
Similarly, some small differences appear in simulation L100, at the very bright end e.g. $M_{1600,\rm \, AB}<-21$. 
The reason for this is that mergers can trigger very strong star formation, thus leading to very high UV radiation.

The effects of supernovae feedback on $\phi$ are only visible at the faint end (e.g. $M_{1600,\rm \, AB}>-18$ of simulation L100 and $M_{1600,\rm \, AB}>-16$ of simulation L35), as supernovae explosions mainly affect star formation in the less massive halos (see the Fig.~\ref{fig:sfr_halo_dis_z7}), that have low SFR and thus low UV luminosity.
Since the fainter galaxies are very hard to detect even for JWST, the impact on $\phi$ caused by supernovae feedback might be hard to confirm with observations of UV luminosity functions. 
Consistently to  Fig.~\ref{fig:sfr_evol} and Fig.~\ref{fig:sfr_halo_dis_z7}, the UV luminosity function $\phi$ is not sensitive to the AGN model. 

The escape fraction parameter (i.e. $\beta_{\rm es}$) dramatically affects the shape of $\phi$, e.g. with $\beta_{\rm es} = -0.5$ both simulations present more faint UV luminosities but fewer bright ones, while with positive $\beta_{\rm es}$ the UV luminosities of massive galaxies are increased, thus the simulations show more bright UV luminosities, but the number of faint ones is reduced.
Compared to the observational results, it seems that they are consistent with $\beta_{\rm es} = 0$.

In summary, the observed UV objects during the EoR are mostly bright ones, their luminosity functions are not very sensitive to the changing of many galaxy formation parameters, thus the UV luminosity function by itself is not enough to constrain the galaxy formation model.
However, some parameters, e.g. the starburst and the escape fraction ones, should be possibly limited by the UV luminosity functions. 

%%%%%%%%%%%%%%%%%%%%%%%%%%%%%%%%%%%%%%%%%%%%%%%%%%
\subsection{Reionization and 21-cm signal}
\label{res:21cm}

\begin{figure*}
\centering
    \includegraphics[width=0.95\linewidth]{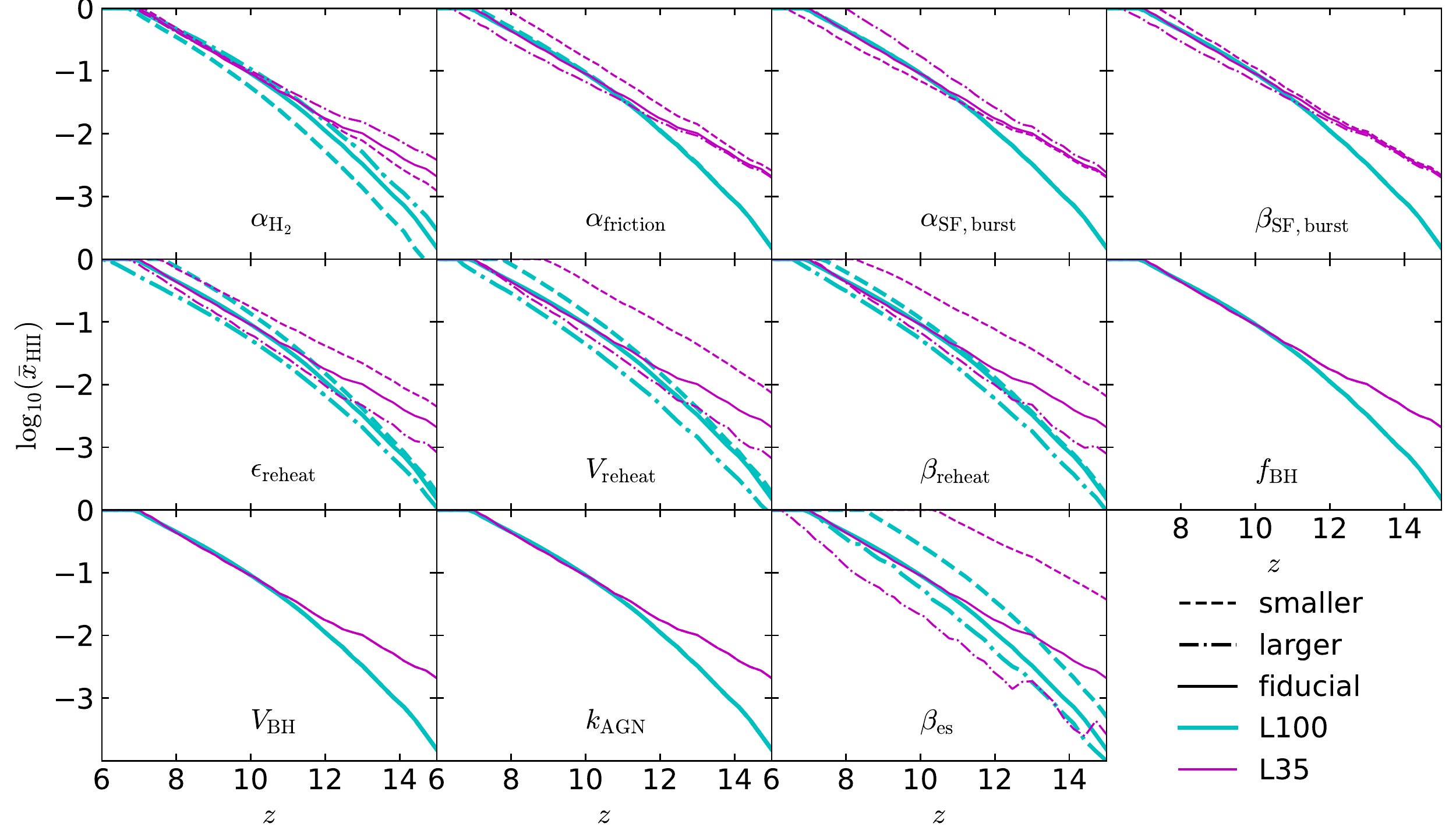}
    \caption{Evolution of the volume averaged ionization fraction ($\bar{x}_{\rm  HII}$) for different values of the parameters $\alpha_{\rm H_{2}}$, $\alpha_{\rm friction}$, $\alpha_{\rm SF,\, burst}$, $\beta_{\rm SF,\, burst}$, $\epsilon_{\rm reheat}$, $V_{\rm reheat}$, $\beta_{\rm reheat}$, $f_{\rm BH}$, $V_{\rm BH}$, $k_{\rm AGN}$ and $\beta_{\rm es}$, from left to right and from top to bottom. 
    Each panel shows the results of the corresponding parameter with smaller value (dashed line), fiducial value (solid line) and larger value (dash-dotted line) in simulation L100 (cyan thick lines) and L35 (magenta thin lines).}
\label{fig:xhii_vz_evol}
\end{figure*}
Fig.~\ref{fig:xhii_vz_evol} shows the history of volume averaged ionization fraction ($\bar{x}_{\rm HII}$) with different galaxy formation and escape fraction model parameters.
Because the number of ionizing photons is related to the stellar mass (see the discussions in Sec.~\ref{meth:rt}), the behaviour of $\bar{x}_{\rm HII}$ looks similar to the one of the stellar mass density (see Fig.~\ref{fig:stellarmass_evol}). 
In the fiducial model $\bar{x}_{\rm HII}$ from L100 is lower than that of L35 at $z>11$ due to the lower SFR density of L100 (see Fig.~\ref{fig:sfr_evol}), while it becomes similar at lower $z$, with $\bar{x}_{\rm HII} = 0.5$ at $z\approx 7.8$. 
Assuming $\bar{x}_{\rm HeII} = \bar{x}_{\rm HII}$ and that helium is fully ionized at $z=3$, the  corresponding CMB optical depth is $\tau \approx 0.059$.
As a reference, the one measured by the {\it Planck} satellite is $0.054 \pm 0.007$ \citep{Planck2020}.

Similarly to the evolution of stellar mass density in Fig.~\ref{fig:stellarmass_evol}, with a smaller (larger) star formation efficiency $\alpha_{\rm H_{2}}$, $\bar{x}_{\rm HII}$ is lower (higher) at $z>12$ in both L100 and L35, while they converge towards the end of the EoR. 
The merger and starburst models affect $\bar{x}_{\rm HII}$ only in L35, with visible differences at $z<10$, when mergers are more frequent. 
For example, a smaller (larger) $\alpha_{\rm friction}$ and $\beta_{\rm SF,\, burst}$ leads to higher (lower) $\bar{x}_{\rm HII}$, while a smaller (larger) $\alpha_{\rm SF,\, burst}$ results in lower (higher) $\bar{x}_{\rm HII}$. 
As supernovae feedback can reduce the SFR and stellar mass density (see Fig.~\ref{fig:sfr_evol} and Fig.~\ref{fig:stellarmass_evol}), it also affects the evolution of $\bar{x}_{\rm HII}$. 
For example, with smaller (larger) $\epsilon_{\rm reheat}$, $V_{\rm reheat}$ and $\beta_{\rm reheat}$, $\bar{x}_{\rm HII}$ throughout the whole EoR period becomes much higher (lower) in both simulations. 
The AGN models do not appreciably affect the ionization process.
As a negative $\beta_{\rm es}$ (i.e. -0.5) increases the budget of ionizing photons from low mass galaxies (stellar mass $M_{\ast}<10^{8}{\rm M}_{\odot}$), it dramatically speeds up the ionization process in both  simulations.
Instead, a positive $\beta_{\rm es}$ (i.e. 0.5) reduces the output of ionization photon radiation from low mass galaxies, while it increases the one from massive galaxies.
However, since there is a paucity of massive galaxies, the net effect is that the positive $\beta_{\rm es}$ delays the ionization process, especially in simulation L35.

\subsubsection{21-cm power spectra at halfway point of EoR}
\begin{figure*}
\centering
    \includegraphics[width=0.95\linewidth]{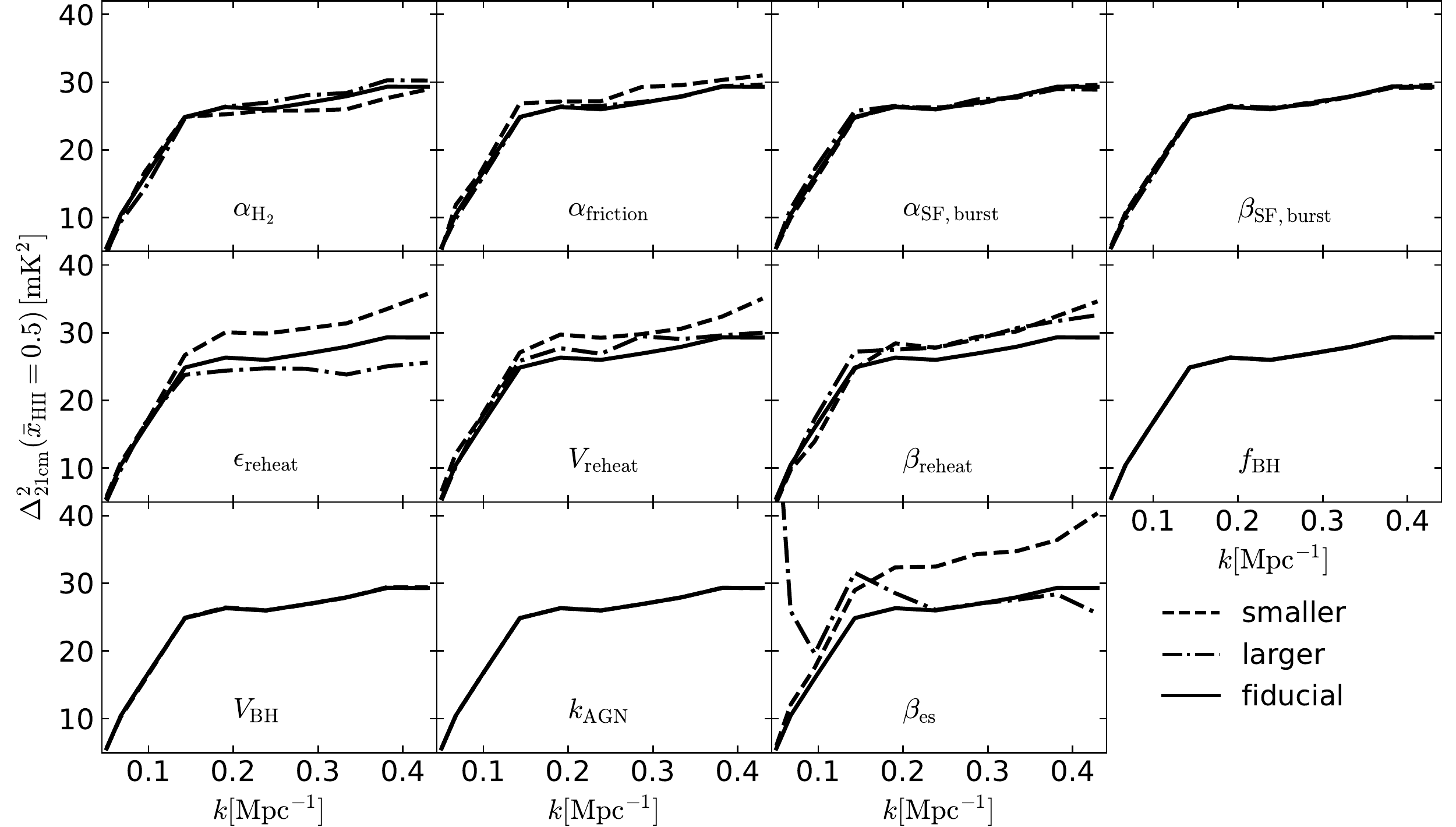}
    \caption{Power spectra $\Delta^2_{\rm 21cm}$ at $\bar{x}_{\rm HII} = 0.5$ with parameters $\alpha_{\rm H_{2}}$, $\alpha_{\rm friction}$, $\alpha_{\rm SF,\, burst}$, $\beta_{\rm SF,\, burst}$, $\epsilon_{\rm reheat}$, $V_{\rm reheat}$, $\beta_{\rm reheat}$, $f_{\rm BH}$, $V_{\rm BH}$, $k_{\rm AGN}$ and $\beta_{\rm es}$, from left to right and from top to bottom. 
    Each panel shows the results of the corresponding parameter with smaller value (dashed line), fiducial value (solid line) and larger value (dash-dotted line) in simulation L100.}
\label{fig:ps_21cm_vk}
\end{figure*}

Fig.~\ref{fig:ps_21cm_vk} shows the power spectra $\Delta^2_{\rm 21cm}(k)$ of the $\delta T_{\rm 21cm}$ at $\bar{x}_{\rm HII} = 0.5$ of simulation L100.
The results from L35 are not shown, as its small cell number (i.e. $35^{3}$) leads to very large sample variance on the power spectra. 
The different values of parameters, e.g. star formation efficiency $\alpha_{\rm H_{2}}$, time delay factor of mergers $\alpha_{\rm friction}$, supernovae feedback models ($\epsilon_{\rm reheat}$, $V_{\rm reheat}$ and $\beta_{\rm reheat}$) and escape fraction index factor $\beta_{\rm es}$, result in obviously different $\Delta^2_{\rm 21cm}$ at $k>0.15\,\rm cMpc^{-1}$, while other parameters -- i.e. starburst model ($\alpha_{\rm SF, \, burst}$ and $\beta_{\rm SF,\, burst}$) and AGN model ($f_{\rm BH}$, $V_{\rm BH}$ and $k_{\rm AGN}$) -- have no significant effects on $\Delta^2_{\rm 21cm}$.

Specifically, a higher (lower) star formation rate speeds up (delays) the ionization process, so that the $\bar{x}_{\rm HII} = 0.5$ value is reached at different redshifts (see Fig.~\ref{fig:xhii_vz_evol}). 
When $\bar{x}_{\rm HII} = 0.5$ happens at higher (lower) redshifts, the $\delta T_{\rm 21cm}$ presents larger (smaller) amplitudes of $\Delta^2_{\rm 21cm}$, especially at small scales.
For example, with a smaller (larger) $\epsilon_{\rm reheat}$ the SFR and stellar mass densities are much higher (lower), thus the ionizing process is faster (slower), with the consequence that $\Delta^2_{\rm 21cm}$ at $k>0.15\,\rm cMpc^{-1}$ is $\sim 10\%$ higher (lower) than in the case with the fiducial $\epsilon_{\rm reheat}$ value. 
A similar effect is associated with the parameters $\alpha_{\rm H_{2}}$ and $\alpha_{\rm friction}$, although the differences induced on $\Delta^2_{\rm 21cm}$ are only few percents.
Differently, both the smaller and larger $V_{\rm reheat}$ and $\beta_{\rm reheat}$ values result in an amplitude of $\Delta^2_{\rm 21cm}$ higher than the fiducial one, due to their complicated relation to the star formation of halos (see Fig.~\ref{fig:sfr_halo_dis_z7}).
Although a larger $V_{\rm reheat}$ and $\beta_{\rm reheat}$ reduce the global SFR and thus delay the ionization process (see Fig.~\ref{fig:sfr_evol} and Fig.~\ref{fig:xhii_vz_evol}), they also increase the SFR of some massive halos (see Fig.~\ref{fig:sfr_halo_dis_z7}), leading to larger size of ionized bubbles around these halos, and thus to higher fluctuations of $\delta T_{\rm 21cm}$, and to a higher $\Delta^2_{\rm 21cm}$. 
With $\beta_{\rm es} = -0.5$, $\bar{x}_{\rm HII} = 0.5$ is obtained at very high $z$ (i.e. $9.3$), which results in a $\Delta^2_{\rm 21cm}$ much higher than the fiducial one.
With $\beta_{\rm es} = 0.5$, the $\Delta^2_{\rm 21cm}$ is similar to the fiducial one at $k>0.2\,\rm Mpc^{-1}$ (i.e. small scales), while much higher than the latter at $k<0.1\,\rm Mpc^{-1}$ (large scales).
It is because the positive $\beta_{\rm es}$ significantly increases the size of the ionized bubbles surrounding very massive galaxies, which in turn changes the fluctuations of $\delta T_{\rm 21cm}$. 

\subsubsection{Evolution of 21-cm power spectra}
\begin{figure*}
\centering
    \includegraphics[width=0.95\linewidth]{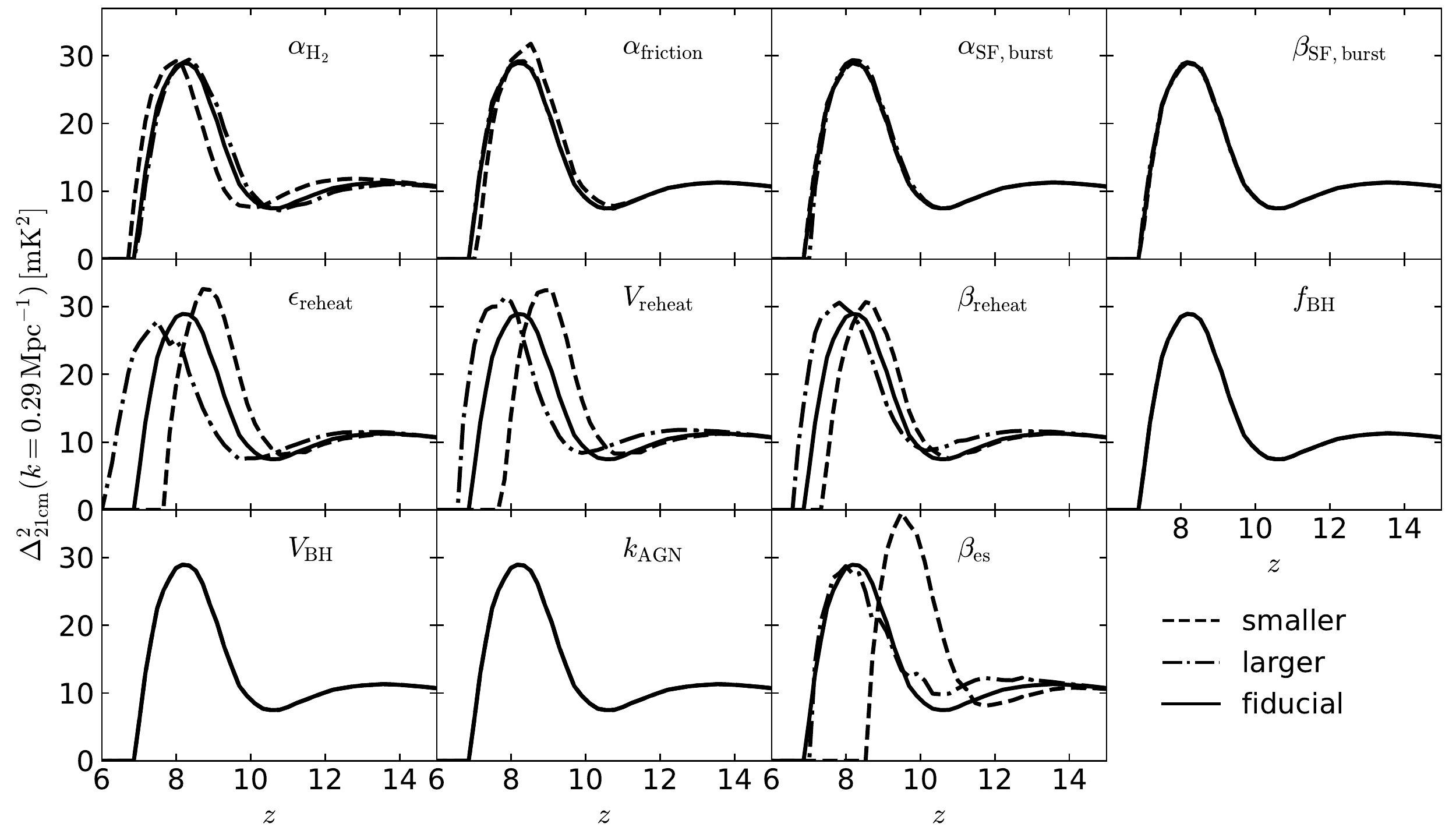}
    \caption{Evolution of 21-cm power spectra $\Delta^2_{\rm 21cm}$ at $k=0.29\,\rm cMpc^{-1}$ with the parameters $\alpha_{\rm H_{2}}$, $\alpha_{\rm friction}$, $\alpha_{\rm SF,\, burst}$, $\beta_{\rm SF,\, burst}$, $\epsilon_{\rm reheat}$, $V_{\rm reheat}$, $\beta_{\rm reheat}$, $f_{\rm BH}$, $V_{\rm BH}$, $k_{\rm AGN}$ and $\beta_{\rm es}$, from left to right and from top to bottom. 
    Each panel shows the results of the corresponding parameter with smaller value (dashed line), fiducial value (solid line) and larger value (dash-dotted line) in the simulation L100.}
\label{fig:ps_21cm_vz}
\end{figure*}
Fig.~\ref{fig:ps_21cm_vz} shows the evolution of $\Delta^2_{\rm 21cm}$ of simulation L100 at $k=0.29\,\rm cMpc^{-1}$, as a function of redshift.
We do not show $\Delta^2_{\rm 21cm}$ at $k=0.1\,\rm cMpc^{-1}$ as it is dominated by sample variance, and thus it is not robust.
Note that the assumption of $T_{\rm S} \gg T_{\rm CMB}$ only works after heating from X-ray sources, thus the results of $\Delta^2_{\rm 21cm}$ are valid only below a certain $z$, which depends on the X-ray source model adopted \citep{Ma2021}.

Since ionization is very weak in the beginning of the EoR, the fluctuations of $\delta T_{\rm 21cm}$ are dominated by the matter density, thus all models present a similar $\Delta^2_{\rm 21cm}$ at $z>13$.
With decreasing redshift, the fluctuations of ionization fraction $x_{\rm HII}$ start to dominate the amplitude of $\Delta^2_{\rm 21cm}$, which peaks at $z \approx 8$ ($\bar{x}_{\rm HII} \approx 0.45$) in the fiducial model.

The differences of $\Delta^2_{\rm 21cm}$ caused by the different values of parameters, are mostly at $z<10$.
Specifically, the supernovae feedback models (i.e. parameters $\epsilon_{\rm reheat}$, $V_{\rm reheat}$ and $\beta_{\rm reheat}$) show the most pronounced differences, as supernovae feedback strongly affects the star formation during the EoR (see Fig.~\ref{fig:sfr_evol}) and thus the ionization history (see Fig.~\ref{fig:xhii_vz_evol}).
Instead, only slight differences are visible for different values of the star formation efficiency $\alpha_{\rm H_{2}}$ and the time delay parameter of merger $\alpha_{\rm friction}$. 
Typically, the higher the redshift at which the peak in $\Delta^2_{\rm 21cm}$ happens, the larger its amplitude is e.g. for parameters $\alpha_{\rm H_{2}}$ and $\epsilon_{\rm reheat}$.
Differently, both the smaller and larger values of $V_{\rm reheat}$ and $\beta_{\rm reheat}$ have peak amplitudes of $\Delta^2_{\rm 21cm}$ higher than the fiducial ones, although their ionization histories are clearly different from each other (see Fig.~\ref{fig:xhii_vz_evol}).
As mentioned earlier, this is due to the complicated dependence of the galactic star formation on $V_{\rm reheat}$ and $\beta_{\rm reheat}$ for different halo masses.
The negative $\beta_{\rm es}$ (i.e. -0.5) leads to very early ionization, and thus to a clearly different $\Delta^2_{\rm 21cm}$ evolution history, while the $\Delta^2_{\rm 21cm}$ with positive $\beta_{\rm es}$ (i.e. 0.5) is roughly consistent with the fiducial one. 
Changing the starburst parameters $\alpha_{\rm SF,\, burst}$ and $\beta_{\rm SF,\, burst}$, as well as the AGN models (i.e. $f_{\rm BH}$, $V_{\rm BH}$ and $k_{\rm AGN}$) does not visibly affect the evolution of $\Delta^2_{\rm 21cm}$.

%%%%%%%%%%%%%%%%%%%%%%%%%%%%%%%%%%%%%%%%%%%%%%%%%%
\section{Discussion and Conclusions}
\label{sec:conc}

Ongoing and upcoming observations, e.g. with the JWST and SKA telescopes, respectively, will enable us to measure both the galaxy properties and the 21-cm signal during the EoR. 
In order to optimally exploit these forthcoming data, we have designed {\sc polar}, a novel semi-numeric algorithm obtained by including the semi-analytical model for galaxy formation {\sc L-Galaxies 2020} \citep{Henriques2020} 
within the 1-D radiative transfer code {\sc grizzly} \citep{Ghara2018}. {\sc polar} is then able to describe consistently both the galaxy formation and the reionization process.
Compared to previous works \cite[e.g. ][]{Park2020, Zhang2022}, our framework is based on a well established and widely used semi-analytic model of galaxy formation, which allows for the inclusion of an extensive network of physical processes. 
{\sc polar} is similar to the semi-numerical models ASTRAEUS \citep{Hutter2021} and MERAXES \citep{Mutch2016}, but with different modelling for galaxy formation and radiative transfer. 
More specifically, while {\sc polar} is based on a 1-D radiative transfer approach which allows also for a more accurate modeling of the source spectra and their effect on the temperature and ionization state of the gas, in ASTRAEUS and MERAXES the evolution of the ionized regions is followed by essentially comparing the number of emitted photons to the number of absorptions. 
While in this paper we only introduce {\sc polar} and explore the effect of a few selected parameters on the galaxy and reionization process, in the future we will use it to perform a parameter fitting based on MCMC techniques, which is possible due to the low computation requirements of {\sc polar}.

With the newly published GADGET-4 code \citep{Springel2021}, we ran two {\it N}-body simulations of limited box length $100\,{\rm cMpc}/h$ (named L100) and $35\,{\rm cMpc}/h$ (named L35), which resolve a minimum halo mass of $\sim 4.2 \times 10^{9}\,\rm M_{\odot}$ and $\sim 1.7 \times 10^{8}\,\rm M_{\odot}$, respectively. 
These simulations have a consistent halo mass function within the range $(4.2 \times 10^{9} - 3.6 \times 10^{11}) \,\rm M_{\odot}$.
Using the merger trees and dark-matter density fields as inputs, and adopting the best-fit values for the galaxy formation parameters from \cite{Henriques2020}, with {\sc polar} we obtain a star formation history, UV luminosity function and CMB Thomson scattering optical depth consistent with observations in the literature.

As this first paper is meant as a proof of concept of our new method, the $N$-body simulations do not reach sizes necessary for 21-cm studies (i.e. several hundreds of cMpc), nor do they resolve small mass halos which could be relevant during the earlier stages of the reionization process. We note that, although {\sc polar} has so far proven to be very efficient, the computation time required to run it on larger or higher resolution simulations will necessarily increase and possibly render an MCMC approach inefficient. 
In this case, we expect to rely on the additional use of specifically designed emulators, similarly to what were done in \cite{Ghara2020, Mondal2022}.
We also note that the inclusion of smaller halos should be accompanied by a modeling of radiative feedback effects, which are expected to affect their star formation (see e.g. \citealt{Hutter2021, Legrand2023}). 

We investigate how the galaxy formation and escape fraction models affect the results in terms of star formation history, UV luminosity function, ionization history and 21-cm power spectrum. 
We find that the star formation and the ionization history are very sensitive to the supernovae feedback models, as supernovae explosion can efficiently reduce star formation within low mass halos. 
They are also significantly affected by the star formation efficiency during the early stage of the EoR, while towards the end of the EoR supernovae feedback can offset the effects of star formation efficiency. 
The starburst triggered by mergers is important in our high resolution simulation L35, while its effects on star formation and ionization are negligible in L100.
The ionization history is very sensitive to the escape fraction model, as it can significantly affect the budget of ionizing photons.
On the contrary, the AGN feedback model does not affect significantly any of the results.

The UV luminosity function is very sensitive to the escape fraction model (e.g. the slope of UV luminosity function), and indeed not all our models are consistent with observations \cite[e.g.][]{Bouwens2021}. 
The parameters describing supernovae feedback and star formation efficiency may be difficult to constrain with observations of the UV luminosity function, as they have an effect only on its faint end, but these faint galaxies are hardly observed.
Differently, since galaxy mergers can trigger very strong star formation and consequently high UV radiation, the merger and starburst model can affect the bright end of the UV luminosity function.

As the 21-cm power spectra from simulation L35 are dominated by sample variance, in this paper we have only discussed those from L100.
We find that they are very sensitive to the supernovae feedback and the escape fraction model, while only weakly sensitive to the star formation efficiency and the galaxy merger model.
Usually, an earlier ionization results in higher amplitudes of the 21-cm power spectra, while we find that both the smaller and larger value of the parameters describing supernovae feedback give a 21-cm power spectrum larger than the one obtained with the fiducial parameter. This is because of their complex dependence on the halo mass. 

{\sc polar}, the new tool introduced in this paper, provides an efficient way build a consistent and realistic galaxy formation and reionization process. In this framework, the different dependence of e.g. UV luminosity functions and 21-cm power spectra on the galaxy formation and escape fraction models would help to reduce the degeneracy between parameters and to exploit at best state-of-the-art multi-wavelength observations from the high redshift universe, as offered by e.g. HST, JWST, ALMA, LOFAR and the planned SKA and EELT (European Extremely Large Telescope).

%%%%%%%%%%%%%%%%%%%%%%%%%%%%%%%%%%%%%%%%%%%%%%%%%%
\section*{Acknowledgements}
The authors would like to thank Rob Yates for his helpful insight into {\sc L-Galaxies 2020}, and an anonymous referee for her/his comments.
QM is supported by the National SKA Program of China (grant No. 2020SKA0110402), National Natural Science Foundation of China (Grant No. 12263002, 11903010), Science and Technology Fund of Guizhou Province (Grant No. [2020]1Y020), and GZNU 2019 Special projects of training new academics and innovation exploration.
RG and SZ acknowledge support grant no. 255/18 from the Israel Science Foundation.
LVEK acknowledges the financial support from the European Research Council (ERC) under the European Union’s Horizon 2020 research and innovation programme (Grant agreement No. 884760, "CoDEX”).
GM acknowledges support by Swedish Research Council grant 2020-04691.
RM is supported by the Israel Academy of Sciences and Humanities \& Council for Higher Education Excellence Fellowship Program for International Postdoctoral Researchers.
The tools for bibliographic research are offered by the NASA Astrophysics Data Systems and by the JSTOR archive.

%%%%%%%%%%%%%%%%%%%%%%%%%%%%%%%%%%%%%%%%%%%%%%%%%%
\section*{Data Availability}
The code {\sc polar}, the simulation data, and also the post-analysis scripts underlying this article will be shared on reasonable request to the corresponding author.



\bibliographystyle{mnras}
\bibliography{ref} % if your bibtex file is called example.bib

%%%%%%%%%%%%%%%%%%%%%%%%%%%%%%%%%%%%%%%%%%%%%%%%%%

%%%%%%%%%%%%%%%%% APPENDICES %%%%%%%%%%%%%%%%%%%%%

\appendix

%%%%%%%%%%%%%%%%%%%%%%%%%%%%%%%%%%%%%%%%%%%%%%%%%%
\section{Independence of iSED on the redshift and model parameters}
\label{app:sed_dep}
\begin{figure}
\centering
    \includegraphics[width=0.95\linewidth]{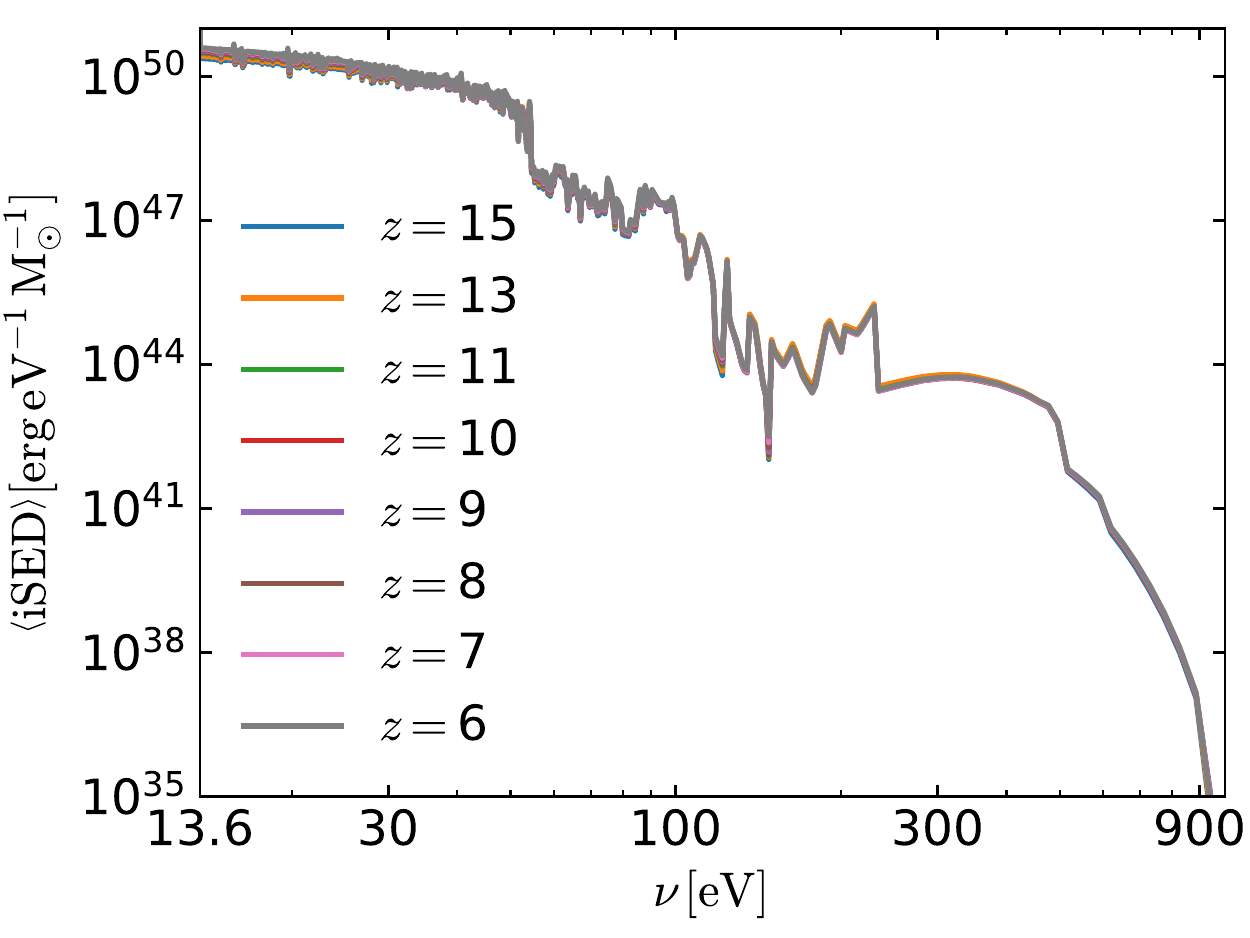}
    \caption{Average iSED ($\rm \left<iSED\right>$) after normalization by the stellar mass of the galaxies at eight redshifts from 15 to 6 in simulation L100 with fiducial parameter values.}
\label{fig:sed_dis_multi_z}
\end{figure}
Fig.~\ref{fig:sed_dis_multi_z} shows the average iSED ($\rm \left<iSED\right>$) from simulation L100 with fiducial parameter values at eight redshifts between 15 and 6.
Note that, to clearly presents the differences caused by the redshift evolution, the 1-$\sigma$ areas are not shown. 
From Fig.~\ref{fig:sed_dis_multi_z} we can see that $\rm \left<iSED\right>$ does not evolve significantly during the EoR. 
This is because, following the integration along cosmic time, the iSED of galaxies is proportional to the stellar mass within galaxies,
so that after normalization by the stellar mass, $\rm \left<iSED\right>$ shows negligible evolution with redshift. 

\begin{figure*}
\centering
    \includegraphics[width=0.95\linewidth]{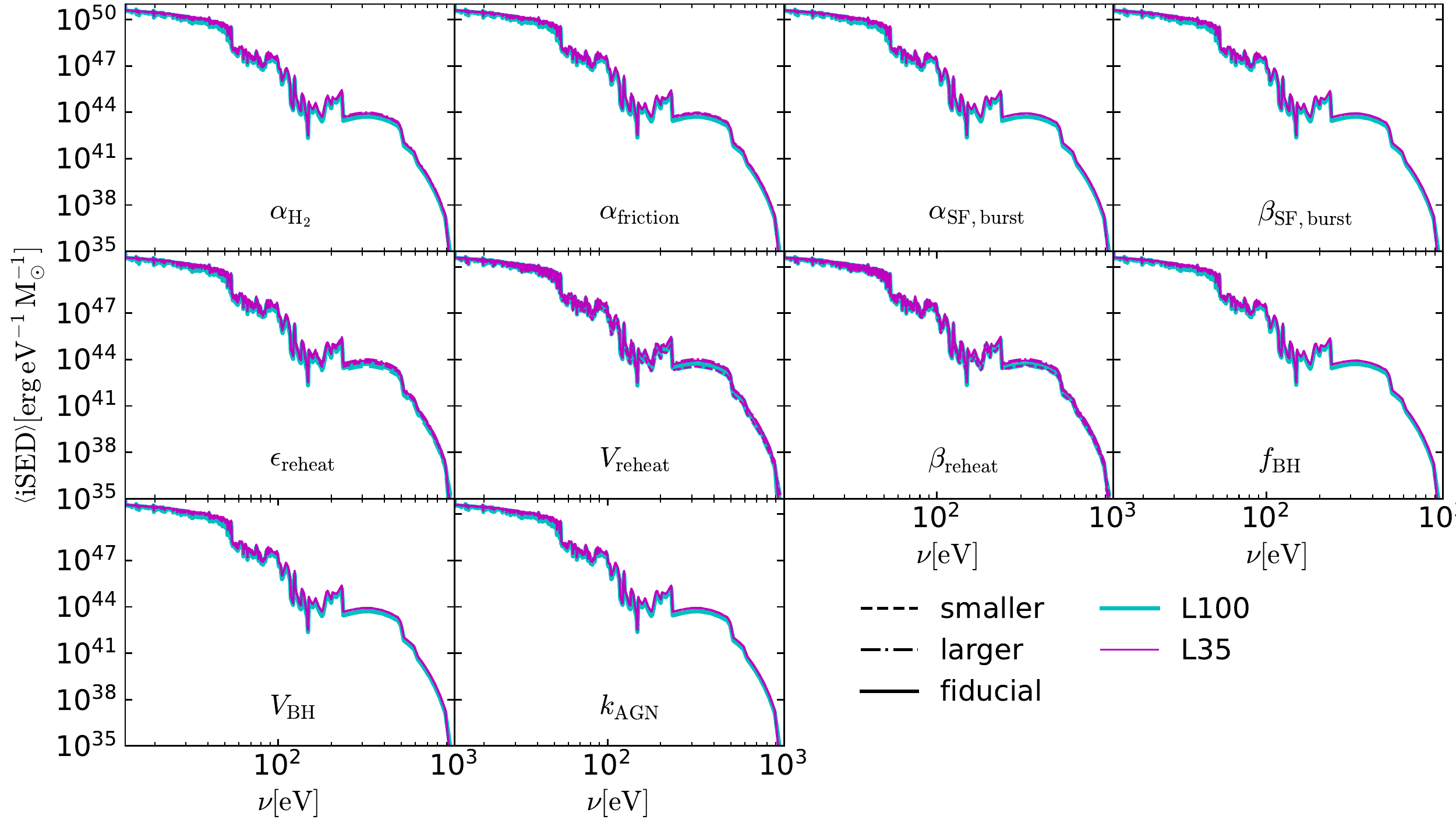}
    \caption{Average iSED ($\rm\left<iSED\right>$) after normalization by the stellar mass of the galaxies at $z=7$, for the galaxy formation parameters $\alpha_{\rm H_{2}}$, $\alpha_{\rm friction}$, $\alpha_{\rm SF,\, burst}$, $\beta_{\rm SF,\, burst}$, $\epsilon_{\rm reheat}$, $V_{\rm reheat}$, $\beta_{\rm reheat}$, $f_{\rm BH}$, $V_{\rm BH}$ and $k_{\rm AGN}$, from left to right and from top to bottom. 
    Each panel shows the results of the corresponding parameter with smaller value (dashed line), fiducial value (solid line) and larger value (dash-dotted line) in the simulation L100 (cyan thick lines) and L35 (magenta thin lines).}
\label{fig:sed_dis_z7_multi}
\end{figure*}
Fig.~\ref{fig:sed_dis_z7_multi} shows $\rm \left<iSED\right>$ at $z=7$ from simulations with different values of the ten free parameters describing the galaxy formation model.
$\rm \left<iSED\right>$ displays no clearly visible changes due to the galaxy formation model, and they are almost the same from simulation L100 and L35. 

%%%%%%%%%%%%%%%%%%%%%%%%%%%%%%%%%%%%%%%%%%%%%%%%%%
\section{Evolution of stellar mass density}
\label{app:stellar_mass}
\begin{figure*}
\centering
    \includegraphics[width=0.95\linewidth]{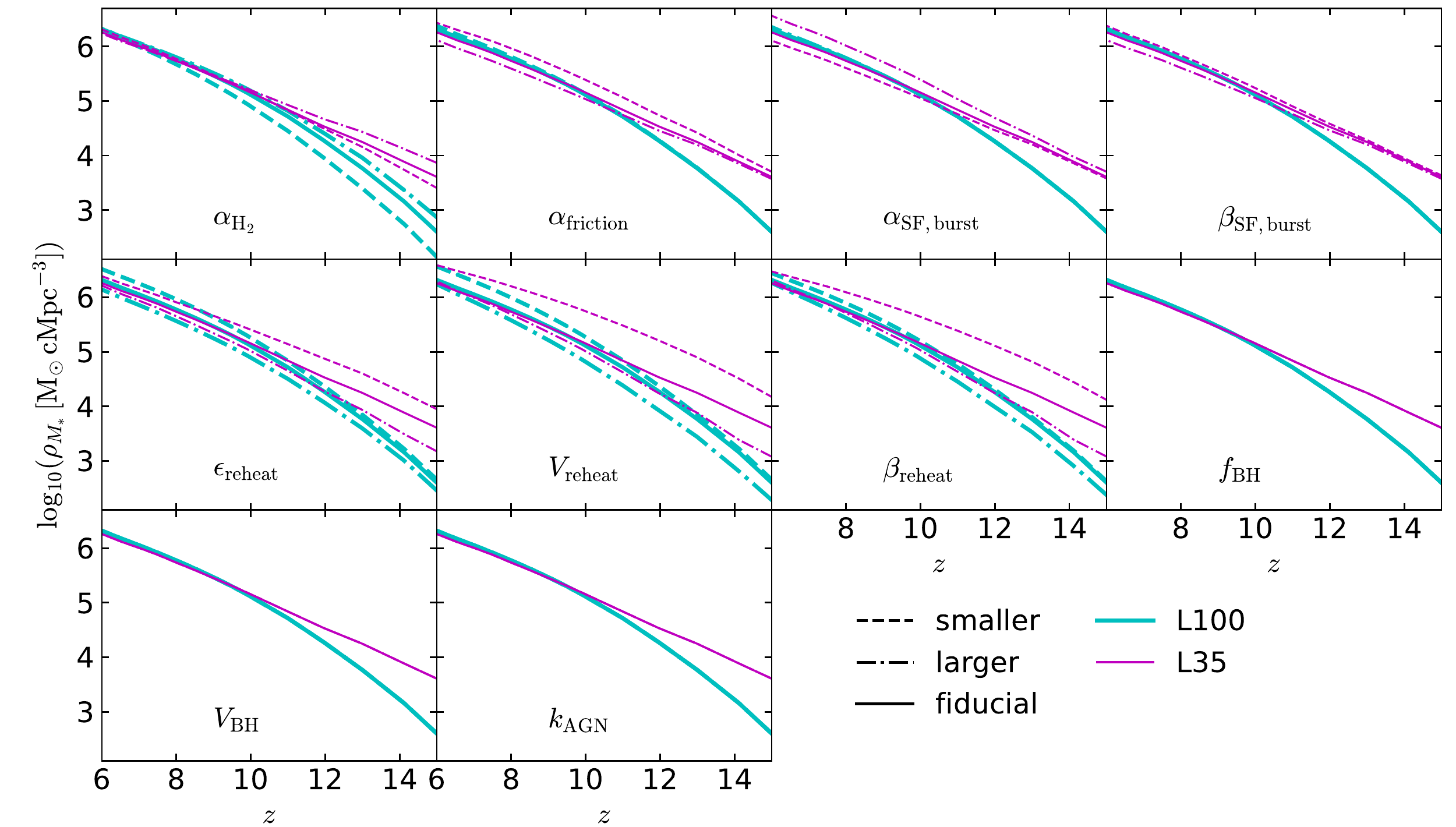}
    \caption{Evolution of stellar mass density $\rho_{M_{\ast}}$ versus redshift $z$ for the galaxy formation parameters $\alpha_{\rm H_{2}}$, $\alpha_{\rm friction}$, $\alpha_{\rm SF,\, burst}$, $\beta_{\rm SF,\, burst}$, $\epsilon_{\rm reheat}$, $V_{\rm reheat}$, $\beta_{\rm reheat}$, $f_{\rm BH}$, $V_{\rm BH}$ and $k_{\rm AGN}$, from left to right and from top to bottom. 
    Each panel shows the results of the corresponding parameter with smaller value (dashed line), fiducial value (solid line) and larger value (dash-dotted line) in the simulation L100 (cyan thick lines) and L35 (magenta thin lines).}
\label{fig:stellarmass_evol}
\end{figure*}
Fig.~\ref{fig:stellarmass_evol} shows the evolution of the stellar mass density $\rho_{M_{\ast}}$ as a function of redshift for different values of the ten free parameters describing the galaxy formation model. 
The evolution features are roughly consistent with those of the SFR shown in Fig.~\ref{fig:sfr_evol}.
Specifically, changing star formation efficiency $\alpha_{\rm H_{2}}$ and supernova feedback parameters (i.e. $\epsilon_{\rm reheat}$, $V_{\rm reheat}$ and $\beta_{\rm reheat}$) visibly affect the evolution of $\rho_{M_{\ast}}$.
The curves corresponding to different values of $\alpha_{\rm H_{2}}$ converge at end of the EoR, due to the supernova feedback that offsets the effects of increasing/decreasing $\alpha_{\rm H_{2}}$.
The impact of starburst due to mergers is significant only in simulation L35, but negligible in L100.
The AGN feedback models have no obvious effects on $\rho_{M_{\ast}}$.

The curves look much smoother than those of the SFR shown in Fig.~\ref{fig:sfr_evol} because the stellar mass within the galaxies results from an integration of the SFR history.

%%%%%%%%%%%%%%%%%%%%%%%%%%%%%%%%%%%%%%%%%%%%%%%%%%
\section{Comparison of UV luminosity functions with observations}
\label{app:uv_lum}
\begin{figure*}
\centering
    \includegraphics[width=0.95\linewidth]{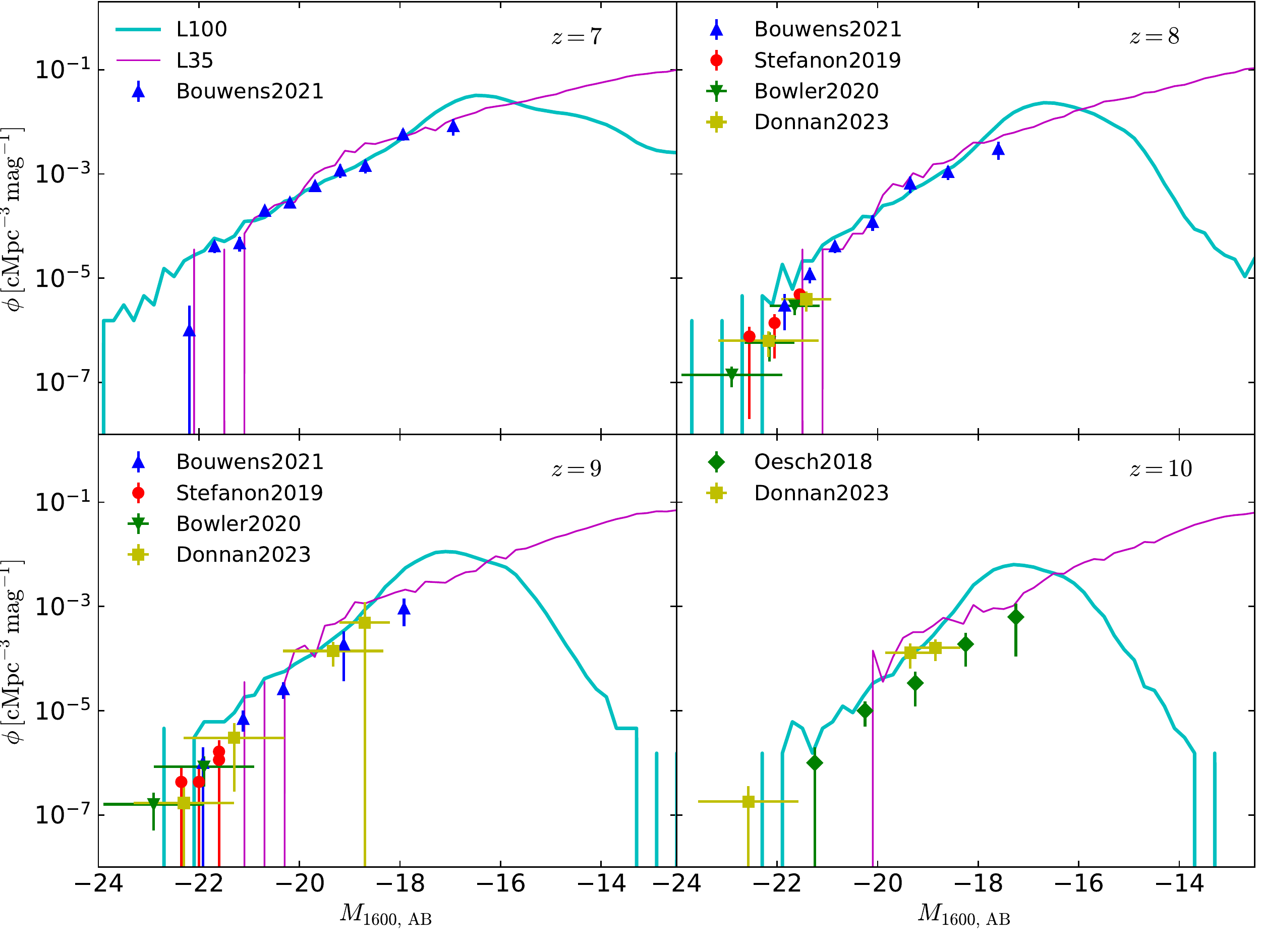}
    \caption{UV luminosity functions $\phi$ at the rest-frame wavelength $\lambda = 1600$ \AA~of the fiducial model in the simulation L100 (cyan thick lines) and L35 (magenta thin lines).
    From left to right, top to bottom, the four panels are the results at $z=7$, 8, 9 and 10, respectively. 
    The observation data points are from \citet{Oesch2018} (green diamond), \citet{Stefanon2019} (red circle), \citet{Bowler2020} (green down triangle), \citet{Bouwens2021} (blue up triangle), and \citet{Donnan2022} (yellow square). 
    Note that the results of \citet{Bowler2020} and \citet{Donnan2022} are at $\lambda = 1500$ \AA.}
\label{fig:uv_lum_multiz}
\end{figure*}
As a supplement to the UV luminosity function $\phi$ shown in Fig.~\ref{fig:lum_uv_dis_z7}, Fig.~\ref{fig:uv_lum_multiz} presents the $\phi$ of the fiducial model at four redshifts in both simulation L100 and L35, together with recent high-$z$ observations from
 HST \citep{Oesch2018, Stefanon2019, Bowler2020, Bouwens2021} and  JWST \citep{Donnan2022}.
The luminosity functions of the fiducial model are broadly consistent with the observations at four redshifts. 
Due to the lack of low mass halos in L100, the corresponding luminosity functions at $M_{\rm 1600,\, AB}>-18$ are not robust, while those from L35 in the same range are consistent with observations. 
Note that the $\phi$ from \citet{Bowler2020} and \citet{Donnan2022} are at $\lambda = 1500$ \AA, while our computed luminosity functions are at $\lambda = 1600$ \AA. However, the differences are expected to be very small.

% Don't change these lines
\bsp	% typesetting comment
\label{lastpage}
\end{document}